\documentclass[a4paper,12pt]{article}
\usepackage{jheppub} 
\usepackage{lineno}

\makeatletter
\gdef\@fpheader{}
\makeatother

\usepackage{mathtools}
\usepackage{amssymb}
\usepackage{amsthm}
\usepackage[mathscr]{eucal}

\usepackage[utf8]{inputenc}
\usepackage{graphicx}
\usepackage{subcaption}
\usepackage[all]{xy}
\usepackage{tensor}
\usepackage{comment}
\usepackage{bm}
\usepackage[normalem]{ulem}
\usepackage[dvipsnames]{xcolor}

\usepackage{ragged2e}
\usepackage{tikz}
\usetikzlibrary{calc}
\usetikzlibrary{patterns}
\usetikzlibrary{decorations.pathmorphing}
\usepackage{physics}
\usepackage{upgreek}
\usetikzlibrary{intersections,patterns}

\hypersetup{
	breaklinks=true,   
	colorlinks=true,   
}

\usepackage{diagbox}  
\usepackage{makecell} 
\usepackage{hhline} 
\usepackage{stackengine}

\usepackage{enumerate}

\usepackage[labelfont={small, bf, sf}, textfont={small,sf}]{caption}
\numberwithin{equation}{section} 

\newcommand{\beq}{\begin{equation}}
	\newcommand{\eeq}{\end{equation}}
\newcommand{\bes}{\begin{subequations}}
	\newcommand{\ees}{\end{subequations}}
\newcommand{\bea}{\begin{eqnarray}}
	\newcommand{\eea}{\end{eqnarray}}

\newcommand{\dL}{\mathcal{L}}
\newcommand{\dG}{\mathcal{G}}

\newtheorem*{theorem-non}{Theorem}

\def\be{\begin{equation}}
	\def\ee{\end{equation}}

\newcount\hh
\newcount\mm
\mm=\time
\hh=\time
\divide\hh by 60
\divide\mm by 60
\multiply\mm by 60
\mm=-\mm
\advance\mm by \time
\def\hhmm{\number\hh:\ifnum\mm<10{}0\fi\number\mm}

\def\be{\begin{equation}}
	\def\ee{\end{equation}}

\numberwithin{equation}{section}

\usepackage{cleveref}
\creflabelformat{equation}{(#2#1#3)}
\crefrangelabelformat{equation}{(#3#1#4)–(#5#2#6)}

\def\O{\mathcal{O}}

\crefname{section}{Section}{Sections}
\crefname{appendix}{Appendix}{Appendices}
\crefname{figure}{Fig.}{Figs.}

\crefname{definition}{Def.}{Defs.}
\crefname{prop}{Prop.}{Props.}
\crefname{lemma}{Lemma}{Lemmas}
\crefname{corollary}{Cor.}{Cors.}
\crefname{thm}{Theorem}{Theorems}
\crefname{remark}{Remark}{Remarks}

\begin{document}
	
\title{The Magnusian generator for dissipative systems
and application to leading 2.5PN radiation-reaction dynamics}
	\date{\today}
	
	\author{Francisco M. Blanco}
	
	\affiliation{Max Planck Institute for Gravitational Physics
(Albert Einstein Institute), Am M\"uhlenberg 1,
14476 Potsdam, Germany}

\emailAdd{francisco.blanco@aei.mpg.de}

	\abstract{The Magnusian is a phase-space function that generates finite-time evolution through nested Poisson brackets. It is related to several familiar generators of classical dynamics, including the radial action, the eikonal phase and related quantities. In this work, we extend the Magnusian framework to systems with dissipation and nonlocal-in-time interactions using the in-in formalism, also known as the Schwinger-Keldysh or Galley formalism. This framework is particularly natural for binary dynamics, where integrating out the mediating gravitational field can produce both dissipative radiation-reaction effects and hereditary, nonlocal-in-time interactions. We derive the generalized Magnusian and show that it continues to generate finite-time evolution. As an application, we construct the Magnusian for Newtonian bound motion subject to the leading 2.5PN radiation-reaction force. The resulting generator defines a discrete evolution map from one cycle to the next and describes the evolution of the system in agreement with numerical solutions.
	}

	\maketitle

\section{Introduction}\label{sec:intro}

Gravitational-wave astronomy has opened a new observational window onto the strong-field regime of gravity. Ground-based interferometers such as LIGO \cite{introLIGO1,introLIGO2,introLIGO3}, together with future space-based observatories such as LISA \cite{Audley:2017drz}, enable precision studies of compact binary sources. Extracting physical information from these signals requires accurate theoretical descriptions of binary dynamics, including both conservative orbital motion and dissipative radiation-reaction effects.

Analytic approaches to the relativistic two-body problem have developed rapidly across post-Newtonian theory \cite{introPN,poissonwill}, effective field theory \cite{Goldberger_2006,Porto:2016pyg,Levi:2018nxp,Dlapa_2022,kalin_2020_third_PM,PRLPorto4PMconservative,porto2026blackholedynamicsfifth}, scattering amplitudes \cite{Damour:2016gwp,Cheung_2018,Vines_2018,Bern_2019,Bern:2021dqo,Bern_2022_PoS,Bern_2022_PRL,bern_2023_fifth_power,general_relativity_from_scattering_amplitudes_2018,PhysRevD.100.084040}, worldline QFT techniques \cite{Mogull_2021wqft,jakobsen2022thingsretardedradiationreactionworldline,conservative5pm1gsf,conservative5pm2gsf,jakobsen_tidal}, gravitational self-force theory \cite{introEMRI,Barack:2018yly,pound,pound3}, and the effective-one-body formalism \cite{intro1body,Damour:2012mv,Taracchini:2013rva}. These frameworks provide complementary perturbative descriptions of compact-binary dynamics across different regimes.

As these calculations are pushed to higher orders, their complexity grows rapidly. It is therefore important to organize perturbative corrections in a way that is efficient, gauge invariant, and directly connected to physical observables. In the relativistic two-body problem, local quantities such as the instantaneous separation or force are generally gauge-dependent. By contrast, many useful observables are defined relationally, by comparing configurations selected by invariant boundary conditions or orbital averages. In scattering problems, quantities like the scattering angle and time delay compare asymptotically incoming and outgoing states \cite{Damour:2016gwp,Gonzo_2025}. In bound systems, invariant quantities such as the redshift are calculated from the accumulated dynamics over a radial period, with the orbit labeled by adiabatic invariants \cite{Detweiler_2008,LeTiec:2011ab,Le_Tiec_2015,Blanchet_2017}. This relational structure has been particularly useful for comparing results obtained in different approximation schemes, for example through first-law relations connecting post-Newtonian calculations, self-force results, and effective-one-body dynamics \cite{LeTiec:2011ab,Le_Tiec_2015,Blanchet_2017}. It is therefore natural to organize the dynamics in terms of finite-time maps, which encode the evolution between the invariantly specified endpoints used to define these observables.

These considerations motivate a closer examination of finite-time evolution maps in phase space. To make this precise, consider a Hamiltonian system on a phase space $\mathcal{P}$ with coordinates $Q^A=(q^\mu,p_\mu)$, and let $H(Q,s)$ denote its Hamiltonian, with $s$ the time variable. The evolution of an observable $\O(Q)$ is governed by
\begin{equation}\label{eq:PoissonHam}
\frac{d\O}{ds}=\{\O,H\}.
\end{equation}
The Hamiltonian generates an infinitesimal canonical flow through phase space. For simplicity, we will restrict the discussion below to observables without explicit $s$-dependence, although the construction extends straightforwardly to the more general case.

The finite-time evolution of any observable is determined by the \emph{Magnusian}. The Magnusian is a phase-space function $\chi(Q,s_f,s_i)$ such that the canonical transformation between the initial and final values of any observable can be written as
\begin{eqnarray}\label{eq:magnusianevolution}
\O_f&=&e^{\{\chi,*\}}\O_i\\
&=&\O_i +\{\chi,\O_i\}+\frac{1}{2}\{\chi,\{\chi,\O_i\}\}+\dots\nonumber
\end{eqnarray}
Thus, while the Hamiltonian generates the evolution infinitesimally, the Magnusian generates the complete finite-time canonical map. 

Structures closely related to the Magnusian have appeared in several areas of classical and quantum dynamics. These include the eikonal or scattering generator \cite{Laenen_2009eikonal,PhysRevD.100.066028eikonal,Di_Vecchia_2021eikonal,DIVECCHIA2024eikonal,Kim:2024grz,Kim:2024svw,Kim:2025hpn}, introduced as the generator mapping incoming scattering data to outgoing data, as well as the eikonal phase of the wavefunction in the WKB approximation \cite{sakurai2017modern}, the classical on-shell action \cite{haddad2025unitarityonshellactionworldline,he2026generalizedunitaritymethodworldline}, and the radial action \cite{Gonzo_2024,Gonzo_2025,K_lin_2020,K_lin_2020II,Cho_2022,Dlapa_2022}. The precise relation among these objects was clarified in Ref.~\cite{Kim:2025gis}.

A simple example illustrates how the Magnusian reproduces the familiar first law of binary dynamics in the case of integrable systems. If the Magnusian depends only on conserved, mutually commuting momenta, $\chi=\chi(p_\mu,s_f,s_i)$, its action on the conjugate coordinates simplifies considerably. Taking $\O=q^\mu$ in Eq.~\eqref{eq:magnusianevolution}, all higher nested Poisson brackets vanish and one obtains
\begin{equation}
q_f^\mu-q_i^\mu=-\frac{\partial\chi}{\partial p_\mu},
\end{equation}
which can be rewritten as
\begin{equation}\label{eq:firstlawmagnusian}
    d(-\chi)=\Delta q^\mu dp_\mu.
\end{equation}
This is the familiar first-law structure appearing in integrable problems: derivatives of a function of the conserved quantities generate changes in the conjugate variables, as in the use of the radial action to obtain orbital periods, periastron advance, and related invariants \cite{LeTiec:2011ab,Le_Tiec_2015,Blanchet_2017}. In such cases, the Magnusian reduces to the radial action and the finite map is generated by a single derivative. The addition of spin or extra field degrees of freedom usually breaks integrability. Then, Eq.~(\ref{eq:firstlawmagnusian}) breaks down. However, the Magnusian continues to generate the complete finite-time canonical map through its exponentiated Poisson-bracket action.

The conventional Magnusian construction assumes Hamiltonian evolution on an ordinary phase space $\mathcal{P}$. After the mediating fields of a binary system have been integrated out, however, the resulting reduced equations of motion generally contain dissipative radiation-reaction forces and hereditary nonlocal-in-time interactions. The corresponding evolution therefore does not define a canonical flow on the ordinary phase space of the two bodies. Recent work managed to incorporate radiative observables into the scattering-generator framework by retaining the radiation field as part of an enlarged Hamiltonian phase space \cite{Kim:2025hpn,gonzo2026}. Here, we pursue a complementary approach in which the field degrees of freedom have already been integrated out and the reduced particle dynamics are described directly.

The purpose of this paper is to extend the Magnusian construction to systems governed by dissipative and potentially nonlocal-in-time equations of motion. We accomplish this using the classical in-in formalism of Galley \cite{Galley1,Galley2}, which is closely related to the Schwinger--Keldysh construction \cite{Schwinger,Keldysh,JordanININ}. In this formalism, the dynamical variables are doubled and reorganized into average and difference variables, denoted by $Q_+^A$ and $Q_-^A$. A force $F_A$ that generates non-Hamiltonian evolution for the physical variables can then be represented as part of a Hamiltonian flow on the doubled phase space. The resulting doubled evolution is canonical, even though its projection onto the physical phase space need not be.

The main result of this paper is that, to first order in the perturbing force, the Magnusian on the doubled phase space is (cf. Eq.~\eqref{eq:magnusiandefinition})
\begin{equation}
\chi_{(1)}(Q_+,Q_-,s_f,s_i)=-Q_-^A\int_{s_i}^{s_f}ds\ M^B{}_A(s,s_i;Q_+)F_B\left[X^{(0)}_{s,s_i}(Q_+)\right].
\end{equation}
Here, $X^{(0)}_{s,s_i}$ denotes the zeroth-order flow and $M^A{}_B$ is its Jacobi propagator. Thus, the force is evaluated along the unperturbed trajectory and transported back to the initial phase-space point by $M^A{}_B$. 

The organization of this work is as follows. In Section \ref{sec:galley}, we review the classical in-in formalism, the doubling of the degrees of freedom, and the way this embeds dissipative dynamics into a Hamiltonian flow on the doubled phase space. In Section \ref{sec:magnusder}, we review the classical interaction picture and derive the relation between the Magnusian and the Hamiltonian. As an application, in Section \ref{sec:2.5} we derive the leading radiation-reaction Magnusian for an eccentric binary governed by Newtonian conservative dynamics supplemented by the 2.5PN dissipative force. The resulting generator produces a discrete evolution map from one cycle to the next and captures the cumulative evolution at leading dissipative order. In Section \ref{sec:nit}, we discuss how the Magnusian formalism relates to near-identity transformations  and explain its difference with the use of NITs in the self-force literature. Finally, in Section \ref{sec:disc}, we summarize our results and discuss possible future applications.

\section{Review of the classical in-in formalism}\label{sec:galley}

In this section, we review the classical in-in formalism introduced in Refs.~\cite{Galley1,Galley2}. We begin with a local conservative system in order to explain the doubling of the degrees of freedom and the physical limit. Dissipative forces and nonlocal-in-time interactions will be introduced subsequently.

Consider a Hamiltonian system with phase-space coordinates $Q^A=(q^\mu,p_\mu)$ and action
\begin{equation}\label{eq:action0}
S[Q]=\int_{s_i}^{s_f}\left[p_\mu\dot q^\mu-H(q,p)\right]ds.
\end{equation}
For fixed initial and final times, the variation of the action is
\begin{equation}\label{eq:single-action-variation}
\delta S=\int_{s_i}^{s_f}\left[\left(\dot q^\mu-\frac{\partial H}{\partial p_\mu}\right)\delta p_\mu-\left(\dot p_\mu+\frac{\partial H}{\partial q^\mu}\right)\delta q^\mu\right]ds+\left[p_\mu\delta q^\mu\right]_{s_i}^{s_f}.
\end{equation}
The bulk terms give Hamilton's equations,
\begin{equation}
\dot q^\mu=\frac{\partial H}{\partial p_\mu},\qquad \dot p_\mu=-\frac{\partial H}{\partial q^\mu}.
\end{equation}
In the standard variational principle, the boundary term in Eq.~\eqref{eq:single-action-variation} is removed by fixing the configuration at both endpoints. The action principle is therefore naturally formulated as a boundary-value problem. For dissipative systems, however, the dynamics are more naturally posed as an initial-value problem: the initial phase-space point is specified, while the final state must be determined by the evolution.

\subsection{Double variational principle and physical limit}
The in-in formalism reformulates the action principle as an initial-value problem by doubling the dynamical variables. One introduces two phase-space histories $Q_1(s)$ and $Q_2(s)$ and defines the doubled action
\begin{equation}\label{eq:doubled-action}
\mathcal{S}[Q_1,Q_2]=S[Q_1]-S[Q_2].
\end{equation}
The two histories are auxiliary variables on an enlarged phase space and do not represent two independent physical systems. The relative minus sign arises from the opposite orientations of the two branches: the first branch runs from $s_i$ to $s_f$, while the second returns from $s_f$ to $s_i$.

On shell, the variation of the doubled action reduces to
\begin{equation}\label{eq:doubled-boundary-variation}
\left.\delta\mathcal{S}\right|_{\mathrm{on-shell}}=\left[p_{1\mu}\delta q_1^\mu-p_{2\mu}\delta q_2^\mu\right]_{s_i}^{s_f}.
\end{equation}

The doubled variational problem is supplemented by the following boundary conditions. At the initial time, the two histories are assigned the same physical initial data,
\begin{equation}\label{eq:initial-equality}
Q_1(s_i)=Q_2(s_i)=Q_i,\qquad \delta Q_1(s_i)=\delta Q_2(s_i)=0.
\end{equation}
At the final time, the two histories are required to coincide, but their common value is left undetermined,
\begin{equation}\label{eq:final-equality}
Q_1(s_f)=Q_2(s_f),\qquad \delta Q_1(s_f)=\delta Q_2(s_f).
\end{equation}
Thus, the final equality condition does not prescribe the final physical state. It only requires the two branches to meet at $s_f$.

With the boundary conditions defined in Eqs.~(\ref{eq:initial-equality}) and (\ref{eq:final-equality}), the initial-time contribution in Eq.~(\ref{eq:doubled-boundary-variation}) vanishes because the initial data are fixed. The final-time contribution vanishes because the values of both paths agree at $s_f$ even if their value is not specified.

The boundary conditions also identify the physical sector of the doubled dynamics. The action in Eq.~\eqref{eq:doubled-action} changes sign under the exchange of the two histories,
\begin{equation}
Q_1\longleftrightarrow Q_2.
\end{equation}
The corresponding equations of motion are therefore mapped into one another by the same exchange. The configuration $Q_1(s)=Q_2(s)$ is consequently a consistent solution of the doubled equations. Since the initial conditions in Eq.~\eqref{eq:initial-equality} assign identical Cauchy data to the two histories, uniqueness of the initial-value problem implies that
\begin{equation}\label{eq:physical-limit}
Q_1(s)=Q_2(s)\equiv Q(s)
\end{equation}
is a valid solution to the doubled dynamics, given appropriate initial data.

Equation~\eqref{eq:physical-limit} defines the \emph{physical limit}. It should not be interpreted as an additional constraint imposed on the equations of motion after they have been derived. Rather, it is the solution selected by choosing physical initial data. Once the two histories begin at the same phase-space point, the exchange symmetry of the doubled equations and uniqueness of the evolution ensure that they remain equal.

Given conservative dynamics determined by an action $S_0[Q]$, dissipation can be incorporated to the doubled phase-space evolution by adding terms that mix the two paths
\begin{equation}
    \mathcal{S}[Q_1,Q_2]=S_0[Q_1]-S_0[Q_2]+K[Q_1,Q_2],
\end{equation}
with $K[Q_1,Q_2]$ antisymmetric under exchange of paths. The dissipative term $K[Q_1,Q_2]$ can be added by hand or it can be obtained after integrating out field degrees of freedom, as we see in Subsection \ref{sec:mediatingfield}.

\subsection{Change of basis to $\pm$ variables}

The \emph{physical limit} becomes particularly transparent in the $\pm$ basis. We define
\begin{equation}\label{eq:pm-definitions}
q_+^\mu=\frac{q_1^\mu+q_2^\mu}{2},\qquad q_-^\mu=q_1^\mu-q_2^\mu,\qquad p_{+\mu}=\frac{p_{1\mu}+p_{2\mu}}{2},\qquad p_{-\mu}=p_{1\mu}-p_{2\mu}.
\end{equation}
Collectively, these variables will be denoted by $Q_+$ and $Q_-$. The average variable $Q_+$ is the average of the two trajectories and reduces to the physical trajectory in the physical limit. The difference variable $Q_-$ measures the separation between the two trajectories and reduces to zero in the physical limit.

The initial equality conditions become
\begin{equation}\label{eq:pm-initial-conditions}
Q_+(s_i)=Q_i,\qquad Q_-(s_i)=0,\qquad \delta Q_+(s_i)=\delta Q_-(s_i)=0.
\end{equation}
The final equality conditions become
\begin{equation}\label{eq:pm-final-conditions}
Q_-(s_f)=0,\qquad \delta Q_-(s_f)=0,
\end{equation}
while $Q_+(s_f)$ remains free. In these variables, the physical limit is the solution
\begin{equation}\label{eq:pm-physical-limit}
Q_+(s)=Q(s),\qquad Q_-(s)=0.
\end{equation}
The condition $Q_-(s)=0$ is not imposed independently at every time. It follows from the physical initial condition $Q_-(s_i)=0$ and the subsequent evolution. Equivalently, the subspace $Q_-=0$ is preserved by the doubled phase-space flow.

The minus sign multiplying the second action in Eq.~\eqref{eq:doubled-action} determines the symplectic structure of the doubled phase space. In the $1,2$ basis, the nonvanishing Poisson brackets are
\begin{equation}\label{eq:12-poisson-brackets}
\{q_1^\mu,p_{1\nu}\}=\delta^\mu{}_\nu,\qquad \{q_2^\mu,p_{2\nu}\}=-\delta^\mu{}_\nu,
\end{equation}
with all brackets between variables belonging to different histories vanishing. In the $\pm$ basis, the nonvanishing brackets are
\begin{equation}\label{eq:pm-poisson-brackets}
\{q_+^\mu,p_{-\nu}\}=\delta^\mu{}_\nu,\qquad \{q_-^\mu,p_{+\nu}\}=\delta^\mu{}_\nu.
\end{equation}
Given the symplectic form $\Omega_{AB}$ on the original phase space, the Poisson bracket between functions $F(Q_+,Q_-)$ and $G(Q_+,Q_-)$ is
\begin{equation}
    \{F,G\}=\Omega^{AB}\left[\frac{\partial F}{\partial Q_+^A}\frac{\partial G}{\partial Q_-^B}+\frac{\partial F}{\partial Q_-^A}\frac{\partial G}{\partial Q_+^B}\right].
\end{equation}

Lastly, we consider the equations of motion in the $\pm$ basis. We expand the doubled action in $\pm$ variables. Since the physical equations are obtained by varying with respect to $Q_-$ and then evaluating at $Q_-=0$, terms beyond linear order in $Q_-$ do not contribute to the physical equations of motion. Hence, we expand to linear order in the difference variables
\begin{equation}\label{eq:doubled-action-expansion}
\mathcal{S}[Q_+,Q_-]=\int_{s_i}^{s_f}Q_-^A(s)E_A[Q_+](s)ds+O(Q_-^2),
\end{equation}
up to boundary terms. Varying with respect to $Q_-$ gives
\begin{equation}\label{eq:variation-minus}
\frac{\delta\mathcal{S}}{\delta Q_-^A(s)}\equiv E_A[Q_+](s)
=0.
\end{equation}
The equation obtained by varying with respect to $Q_+$ is
\begin{equation}\label{eq:variation-plus}
\frac{\delta\mathcal{S}}{\delta Q_+^A(s)}=\int_{s_i}^{s_f}\frac{\delta E_B[Q_+](s')}{\delta Q_+^A(s)}Q_-^B(s')ds'=0.
\end{equation}
The equations obtained from varying with respect to $Q_+$ are the linearized equation governing deviations from the physical trajectory. The difference variables are therefore linearized deviations along the physical path. To make this interpretation explicit, let the physical equations be written as a first-order flow,
\begin{equation}
\dot Q_+^A=V^A(Q_+,s).
\end{equation}
Linear perturbations to this equation satisfy
\begin{equation}\label{eq:Q-}
\dot Q_-^A(s)=\left.\frac{\partial V^A(Q,s)}{\partial Q^B}\right|_{Q=Q_+(s)}Q_-^B(s).
\end{equation}
Let $X_{s,s_i}(Q_i)$ denote the solution at time $s$ whose initial value at $s_i$ is $Q_i$,
\begin{equation}
Q_+^A(s)=X^A_{s,s_i}(Q_i).
\end{equation}
We define the Jacobi propagator\footnote{This is the phase-space equivalent of the Jacobi propagator as defined by Dixon, which describes the linearized evolution of geodesic deviations.} \cite{jacdixon1,jacdixon2,jacdixon3} by
\begin{equation}\label{eq:defjacobi}
M^A{}_B(s,s_i;Q_i)\equiv\frac{\partial X^A_{s,s_i}(Q_i)}{\partial Q_i^B}.
\end{equation}
It satisfies the matrix differential equation
\begin{equation}
\frac{d}{ds}M^A{}_B(s,s_i;Q_i)=\left.\frac{\partial V^A(Q,s)}{\partial Q^C}\right|_{Q=X_{s,s_i}(Q_i)}M^C{}_B(s,s_i;Q_i),
\end{equation}
together with the initial condition
\begin{equation}
M^A{}_B(s_i,s_i;Q_i)=\delta^A{}_B.
\end{equation}
The solution of the linearized equation \eqref{eq:Q-} can therefore be written as
\begin{equation}\label{eq:evolutionQ-}
Q_-^A(s)=M^A{}_B(s,s_i;Q_i)Q_-^B(s_i).
\end{equation}
Physical initial data impose $Q_-^A(s_i)=0$, and Eq.~\eqref{eq:evolutionQ-} then guarantees that $Q_-^A(s)=0$ at all later times. Nevertheless, $Q_-$ is retained as an auxiliary coordinate on the doubled phase space while constructing the evolution generator; the physical limit is taken only after the relevant variations or Poisson brackets have been evaluated. For geodesic motion, Eq.~\eqref{eq:defjacobi} becomes the geodesic-deviation equation in phase space.

\subsection{Effective dynamics from integrating out a mediating field}\label{sec:mediatingfield}

We now consider a point particle of mass $m$ with phase-space coordinate $Q^A$ interacting with some field $\Phi_a$. The field $\Phi_a$ could be a scalar field $\phi$, an electromagnetic field $A_\mu$, or the metric tensor $g_{\mu\nu}$. We are interested in the effective action functional obtained by integrating out the mediating field $\Phi_a$. The action functional for a single copy of the system is
\begin{equation}
    S[Q,\Phi]=S_{\text{m}}[Q]+S_{\text{f}} [\Phi]+S_{\text{int}}[Q,\Phi].
\end{equation}
The equations of motion are 
\begin{subequations}
    \begin{eqnarray}
        \frac{\delta S_{\text{m}}}{\delta Q^A}&=&-\frac{\delta S_{\text{int}}}{\delta Q^A},\\
    \label{eq:fieldeqsaction}    \frac{\delta S_{\text{f}}}{\delta \Phi_a}&=&-\frac{\delta S_{\text{int}}}{\delta \Phi_a}.
    \end{eqnarray}
\end{subequations}
It will be convenient to define the force
\begin{equation}
    F_A[Q]\equiv -\left.\frac{\delta S_{\text{int}}}{\delta Q^A}\right|_{\Phi \text{ on-shell}}.
\end{equation}
Now, consider the action functional with double degrees of freedom
\begin{equation}
    \mathcal{S}=S_1[Q_1,\Phi_1]-S_2[Q_2,\Phi_2].
\end{equation}
We expand this action functional to linear order in the difference variables, defined analogously to Eq.~(\ref{eq:pm-definitions}). The matter action functional is given by 
\begin{equation}
S_\text{m}[Q_1]-S_\text{m}[Q_2]=\int_{s_i}^{s_f}\Big[p_{+\mu}\dot{q}^\mu_-+p_{-\mu}\dot{q}_+^\mu-\mathcal{H}_0(Q_+,Q_-)\Big]ds,
\end{equation}
where the Hamiltonian is expanded to linear order in $Q_-$. The remaining terms are
\begin{subequations}
\begin{eqnarray}
S_{\text{f}}[\Phi_1]-S_{\text{f}}[\Phi_2] &=& \int d^4x \left.\frac{\delta S_\text{f}}{\delta \Phi_a}\right|_{\Phi_+}\Phi_{-a},\label{eq:fieldgalley1}\\
S_{\text{int}}[Q_1,\Phi_1]-S_{\text{int}}[Q_2,\Phi_2]&=&  \int d^4x \left.\frac{\delta S_\text{int}}{\delta \Phi_a}\right|_{Q_+,\Phi_+}\Phi_{-a}+\int ds \left.\frac{\delta S_\text{int}}{\delta Q^A}\right|_{Q_+,\Phi_+} Q_-^A  .\label{eq:intgalley1}  \ \ \ \ \ \ \ \ \ \ 
\end{eqnarray}
\end{subequations}
Now, we evaluate the field $\Phi_a$ on the solution to eq. (\ref{eq:fieldeqsaction}) with the appropriate boundary conditions\footnote{The in-in boundary conditions force the use of the retarded Green's function to solve the field equations. By contrast, the use of endpoint boundary conditions, as is typical for variational principles, leads to time-even Green's functions. This is one of the reasons why the doubled phase-space is suited to include dissipation. We discuss this in more detail in Appendix \ref{appendixscalar}.}. Once the field $\Phi_a$ is evaluated on-shell, the first term in Eq. (\ref{eq:fieldgalley1}) cancels with the first term in Eq. (\ref{eq:intgalley1}), leaving
\begin{equation}\label{eq:generalgalleyaction}
    \mathcal{S}[Q_+,Q_-]=\int_{s_i}^{s_f}\big[p_{+\mu}\dot{q}_-^\mu+p_{-\mu}\dot{q}_+^\mu-\mathcal{H}_0(Q_+,Q_-)\big]ds-\int_{s_i}^{s_f} ds F_A[Q_+]Q_-^A(s).
\end{equation}
We find that solving the doubled field equations with the in-in boundary conditions and substituting the solutions back into the action produces an effective interaction between the two paths, which to leading order in $Q_-$ is represented by the last term in Eq.~(\ref{eq:generalgalleyaction}). The effective Hamiltonian is
\begin{equation}\label{eq:doubledhamforce}
    \mathcal{H}(Q_+,Q_-)=\mathcal{H}_0(Q_+,Q_-)+Q_-^AF_A(Q_+).
\end{equation}
The force $F_A$ can be derived by integrating out the mediating field, as we illustrate in Appendix~\ref{appendixscalar}, or obtained directly from the equations of motion, as we do in Section~\ref{sec:2.5} for the 2.5PN radiation-reaction force.

We note that the force can, in principle, be nonlocal in $Q_+$. This means that it is not a function of the instantaneous value of $Q_+$, but a functional of its entire history $Q_+(s)$. The in-in boundary conditions then select retarded causality for the on-shell fields. This case is studied in detail in Appendix \ref{appendixnonlocal}, where we show that the Magnusian can still be constructed after the nonlocalities have been treated perturbatively.

\subsection{Closure under the doubled Poisson bracket}\label{sec:closure}

As explained in the previous subsections, only linear terms in the difference variables $Q_-$ are required to derive the full dynamics. Now, we show that the class of functions linear in $Q_-$ is closed by the doubled Poisson bracket and that the action by the Poisson bracket is a representation of the action of vector fields on the original phase space. 

Consider two functions of the form
\begin{equation}\label{eq:K1K2}
K_1(Q_+,Q_-)=Q_-^A a_A(Q_+),
\qquad
K_2(Q_+,Q_-)=Q_-^A b_A(Q_+).
\end{equation}
Denoting the symplectic form on the ordinary phase space by $\Omega_{AB}$, the Poisson bracket on the doubled phase space can be written as
\begin{equation}
\{F,G\}=\Omega^{AB}
\left(\frac{\partial F}{\partial Q_+^A}
\frac{\partial G}{\partial Q_-^B}
+\frac{\partial F}{\partial Q_-^A} \frac{\partial G}{\partial Q_+^B}
\right).
\end{equation}
Substituting Eq.~(\ref{eq:K1K2}) into the doubled Poisson bracket gives
\begin{equation}
\{K_1,K_2\}=Q_-^C\,\Omega^{AB}
\left(\frac{\partial a_C}{\partial Q_+^A}b_B
+
a_A\frac{\partial b_C}{\partial Q_+^B}
\right).
\end{equation}
Defining
\begin{equation}
c_C(Q_+)\equiv\Omega^{AB}\left(\frac{\partial a_C}{\partial Q_+^A}b_B+a_A\frac{\partial b_C}{\partial Q_+^B}\right),
\end{equation}
we obtain
\begin{equation}
\{K_1,K_2\}=Q_-^C c_C(Q_+).
\end{equation}
The Poisson bracket is therefore again linear in the difference variables. Hence, the space of functions of the form
\begin{equation}\label{eq:linearin-}
K(Q_+,Q_-)=Q_-^A k_A(Q_+)
\end{equation}
is closed under the doubled Poisson bracket.

Given a function of the form in Eq.~(\ref{eq:linearin-}), its Poisson bracket with a physical observable $\mathcal{O}(Q_+)$ acts as a phase-space vector field
\begin{equation}
    \{K,\mathcal{O}\}=\Omega^{AB}k_A\partial_B\mathcal{O}(Q_+).
\end{equation}
We conclude that the action of functions linear in $Q_-$ under the Poisson bracket provides a representation of the action of vector fields on the physical phase space. Furthermore, the Poisson bracket of a linear function on $Q_-$ with a function of $Q_+$ is always independent of $Q_-$.

\section{Construction of the Magnusian in the interaction picture}\label{sec:magnusder}

In this section, we explain the relation between the Magnusian $\chi$ and the Hamiltonian $H$ and derive an expression relating the two. This relation is particularly simple when the Hamiltonian is time-independent. In that case, the evolution equation
\begin{equation}
    \frac{d\O}{ds}=\{\O,H\}
\end{equation}
can be straightforwardly exponentiated to
\begin{equation}\label{eq:exptimindham}
    \O_f=e^{-(s_f-s_i)\{H,*\}}\O_i.
\end{equation}
By comparison with Eq. (\ref{eq:magnusianevolution}), the Magnusian in this case is simply
\begin{equation}
\chi(Q,s_f,s_i)=-H(Q)(s_f-s_i)   
\end{equation}
Although exact, Eq. (\ref{eq:exptimindham}) is generally not useful as a perturbative expansion over finite time intervals: the Hamiltonian contains the full, order-unity dynamics, and the exponent need not be small. For a perturbed system, this difficulty can be avoided by passing to the interaction picture, where the unperturbed evolution is treated exactly and the transformed Hamiltonian is proportional to the perturbation. The interaction-picture Hamiltonian is therefore perturbatively small and the exponential can be truncated to a finite order. However, interaction-picture Hamiltonians are generally time-dependent, even when the original Hamiltonian is not. In that case, Eq. (\ref{eq:exptimindham}) no longer holds. The relation between the Magnusian and the Hamiltonian is then governed by the Magnus expansion \cite{magnusseries}. In this section, we develop this construction and derive the resulting relation between the interaction-picture Hamiltonian and the Magnusian.

\subsection{The classical interaction picture}\label{sec:interaction}

We begin by defining the interaction picture directly in terms of phase-space flows, following Ref. \cite{kim2025classicaleikonalrelativisticscattering}.

Consider a Hamiltonian of the form
\begin{equation}
H(Q,s)=H_0(Q,s)+\epsilon V(Q,s),
\end{equation}
where the dynamics generated by $H_0$ are assumed to be exactly solvable and $\epsilon V$ is treated perturbatively. Let $X_{s_f,s_i}$ denote the full Hamiltonian flow,
\begin{equation}
\frac{d}{ds_f}X^A_{s_f,s_i}(Q)=\left.\{Q^A,H\}\right|_{Q=X_{s_f,s_i}(Q)},
\end{equation}
and let $X^{(0)}_{s_f,s_i}$ denote the flow generated by $H_0$,
\begin{equation}
\frac{d}{ds_f}X^{(0)A}_{s_f,s_i}(Q)=\left.\{Q^A,H_0\}\right|_{Q=X^{(0)}_{s_f,s_i}(Q)}.
\end{equation}
The interaction-picture flow can be defined as the composition
\begin{equation}
X^I_{s_f,s_i}=X^{(0)}_{s_i,s_f}\circ X_{s_f,s_i}.
\end{equation}
Starting from an initial condition $Q_i$ at time $s_i$, this map first evolves the system forward with the full dynamics to time $s_f$ and then transports the resulting endpoint backward with the unperturbed flow. It therefore acts as a comparison map between the full and zeroth-order trajectories: the point $X^I_{s_f,s_i}(Q_i)$ is the initial condition whose unperturbed evolution reaches the same endpoint that $Q_i$ would reach under the full dynamics.

To make this relation explicit, define the perturbed and unperturbed final-time observables by
\begin{equation}
\O_f(Q_i)\equiv\O\big[X_{s_f,s_i}(Q_i)\big],\qquad \O_f^{(0)}(Q_i)\equiv\O\big[X^{(0)}_{s_f,s_i}(Q_i)\big].
\end{equation}
The definition of the interaction-picture map immediately implies
\begin{equation}
\O_f(Q_i)=\O_f^{(0)}\big[X^I_{s_f,s_i}(Q_i)\big].
\end{equation}
Because $X^I_{s_f,s_i}$ is a composition of canonical transformations, it is itself canonical and is generated by an interaction-picture Magnusian $\chi^I$,
\begin{equation}\label{eq:magnusianevolutioninteraction}
\O_f=e^{\{\chi^I,*\}}\O_f^{(0)}.
\end{equation}
The interaction-picture Magnusian therefore packages all perturbative corrections to the solvable dynamics into a single phase-space function. It describes not the full motion itself, but the finite canonical transformation that relates the unperturbed and perturbed evolutions.

\subsection{The interaction-picture Hamiltonian on the double phase-space}

The preceding construction defines the interaction picture as a comparison between phase-space flows. To obtain its Hamiltonian generator in the doubled formalism, we now represent the same transformation as a time-dependent canonical transformation.

Let
\begin{equation}
\Phi:\mathcal{P}\rightarrow\mathcal{P}
\end{equation}
be a canonical transformation acting on the physical phase-space coordinates according to
\begin{equation}
Q'=\Phi(Q).
\end{equation}
To linear order in the difference variables, the induced transformation on the doubled phase space is
\begin{equation} Q_+'^A = \Phi^A(Q_+), \qquad Q_-'^A = \frac{\partial\Phi^A(Q_+)} {\partial Q_+^B} Q_-^B. \end{equation} 
Thus, $Q_+^A$ transforms as a phase-space coordinate while $Q_-^A$ transforms as a tangent vector at $Q_+$.

For the next few equations, we suppress the $\pm$ labels to avoid clutter, but all functions and functionals depend on both $Q_+$ and $Q_-$. Locally, we represent this canonical transformation by a type-1 generating function $B(s_i,q^I,q)$. Recall that a canonical transformation changes the phase-space action by at most a boundary term
\begin{equation}\label{eq:equateactionscanonical}
    \mathcal{S}^I[Q^I]=\mathcal{S}[Q]+B(s_f,q_f^I,q_f)-B(s_i,q_i^I,q_i)
\end{equation}
The boundary term is easily rewritten as
\begin{equation}\label{eq:expandboundaryterm}
\begin{aligned}        
  B(s_f,q^I_f,q_f)-B(s_i,q^I_i,q_i)&=  \int_{s_i}^{s_f}\frac{d}{ds}B\big(s,q^I(s),q(s)\big)ds\\
  &=\int_{s_i}^{s_f}\left[\frac{\partial B}{\partial s}+\frac{\partial B}{\partial q^{I\mu}}\dot{q}^{I\mu}+\frac{\partial B}{\partial q^\mu}\dot{q}^\mu\right]ds
\end{aligned} 
\end{equation}
Substituting Eq.~(\ref{eq:expandboundaryterm}) into Eq.~(\ref{eq:equateactionscanonical}) we get
\begin{subequations}
    \begin{eqnarray}
        H^I(Q^I)&=&H(Q)-\frac{\partial B}{\partial s}\\
        p_\mu^I&=&\frac{\partial B}{\partial q^{I\mu}}\\
        p_\mu &=&-\frac{\partial B}{\partial q^{\mu}}
    \end{eqnarray}
\end{subequations}
Now, we choose $B(s,q^I,q)$ to be the zeroth-order on-shell action 
\begin{equation}
    B(s,q^I,q,s_i)=-\int_{s_i}^{s} \left[p_\mu^{(0)}\dot{q}^{(0)\mu}-H_0(Q^{(0)})\right]ds'.
\end{equation}
Note that $B$ actually depends on $s$ and on $s_i$. This is not a problem since $s_i$ is a fixed parameter that does not evolve with $s$. In particular, this means that only the boundary term evaluated at $s_f$ survives, since $B$ evaluated at $s=s_i$ vanishes.
The zeroth-order on-shell action satisfies
\begin{subequations}
    \begin{eqnarray}
        \frac{\partial B}{\partial s}&=&H_0(Q)\\
        \frac{\partial B}{\partial q^{I\mu}}&=&p^{(0)}_\mu(s_i)\\
        \frac{\partial B}{\partial q^\mu}&=&-p^{(0)}_\mu(s)
    \end{eqnarray}
\end{subequations}
Hence, the new Hamiltonian is 
\begin{equation}
    H^I(Q^I)=H(Q)-H_0(Q)
\end{equation}
where the right-hand side is understood to be evaluated on $Q(Q^I)$. The perturbed Hamiltonian is given by Eq.~(\ref{eq:doubledhamforce}). The variables are related via
\begin{equation}
    Q^A(s)=X^{(0)}_{s,s_i}\big(Q^I(s)\big).
\end{equation}
Recall that the zeroth-order solution for the difference variables is given by Eq.~(\ref{eq:evolutionQ-})
\begin{equation}
    Q_-^{(0)A}(s)=M^A{}_B(s,s_i;Q_+)Q^B_-(s_i).
\end{equation}
In the remainder of the paper, we suppress the superscript $I$ on the interaction-picture variables. The interaction-picture Hamiltonian is then
\begin{equation}\label{eq:interactionham}
H^I(Q_+,Q_-,s,s_i)=F_A\big[X^{(0)}_{s,s_i}(Q_+)\big]M^A{}_B(s,s_i;Q_+) Q_-^B.
\end{equation}
Note that the force is evaluated along the unperturbed trajectory and pulled back by the Jacobi propagator so that all indices are contracted at the same phase-space location.

\subsection{Derivation of the Magnusian}

We now derive the relation between the Magnusian and the Hamiltonian in the general case where the latter is time-dependent. This derivation was given in Ref.~\cite{Kim:2025gis} for arbitrary Hamiltonian systems; we reproduce it here for clarity.

In the doubled formalism, the Magnusian depends on the average and difference variables and on the two endpoint times,
\begin{equation}
    \chi=\chi(Q_+,Q_-,s_f,s_i)
\end{equation}
The evolution of any observable $\O(Q_+)$ is given by
\begin{equation}\label{eq:Magnusianevolv}
    \O_f=e^{\{\chi,*\}}\O_i
\end{equation}

For the moment, let $\mathcal{H}(Q_+,Q_-,s)$  be an arbitrary time-dependent Hamiltonian within the class  of linear functions in $Q_-$ relevant to the doubled dynamics. The instantaneous evolution of an observable $\O(Q_+)$ is
\begin{equation}\label{eq:poissonHam2}
\frac{d \O}{ds}=\left\{\O, \mathcal{H}\right\},
\end{equation}
We define the operator $N = \{\chi,*\}$ for short. Taking a time derivative of Eq. (\ref{eq:Magnusianevolv}), we get
\begin{eqnarray}\label{eq:timederofmagnop}
    \frac{d\O_f}{ds_f}&=&\frac{d}{ds_f} e^N \O_i\nonumber\\
    &=& \left(\frac{d}{ds_f}e^N\right)  e^{-N}\O_f.
\end{eqnarray}
It follows from Eq. (\ref{eq:timederofmagnop}) and Eq. (\ref{eq:poissonHam2}) that
\begin{equation}
   \left(\frac{d}{ds_f} e^N\right) e^{-N}=\{-\mathcal{H}_f,*\},
\end{equation}
where $\mathcal{H}_f$ denotes the Hamiltonian at final time. Next, we use the identity
\begin{equation}
     \left(\frac{d}{ds_f} e^N\right) e^{-N}=\frac{e^{[N,*]}-1}{[N,*]}\frac{dN}{ds_f}
\end{equation}
to get
\begin{equation}\label{eq:stepbrackets}
\frac{e^{[N,*]}-1}{[N,*]}\frac{dN}{ds_f}=\{-\mathcal{H}_f,*\}.    
\end{equation}
The commutator of the action of Poisson brackets satisfies the following property
\begin{equation}
    [\{A,*\},\{B,*\}]=\{\{A,B\},*\}.
\end{equation}
Hence, the left-hand side of Eq.~(\ref{eq:stepbrackets}) becomes
\begin{equation}
    \frac{e^{[N,*]}-1}{[N,*]}\frac{dN}{ds_f}=\Big\{\frac{e^{N}-1}{N}\frac{d\chi}{ds_f},*\Big\}.
\end{equation}
Relating this back to Eq.~(\ref{eq:stepbrackets}), we get
\begin{equation}
   \frac{e^{N}-1}{N}\frac{d\chi}{ds_f}=-\mathcal{H}_f .
\end{equation}
The operator on the left-hand side can be formally inverted to obtain
\begin{equation}
    \frac{d\chi}{ds_f}=-\frac{\{\chi,*\}}{e^{\{\chi,*\}}-1}\mathcal{H}_f.
\end{equation}
Lastly, switching from total to partial derivative by using
\begin{equation}
    \frac{d\chi}{ds_f}=\frac{\partial \chi}{\partial s_f} + \{\chi,\mathcal{H}_f\},
\end{equation}
we get
\begin{equation}\label{eq:magnusianrelationtoH}
\frac{\partial \chi(Q_+,Q_-,s_f,s_i)}{\partial s_f}=-\frac{\left\{*,\chi\right\}}{e^{\{*,\chi\}}-1}\mathcal{H}(Q_+,Q_-,s_f),
\end{equation}
which can be solved along with the condition $\chi(Q_+,Q_-,s_i,s_i)=0$.

\subsection{Magnus series and perturbative expansion}

If we apply Eq. (\ref{eq:magnusianrelationtoH}) to the interaction-picture Hamiltonian, its right-hand side can be expanded perturbatively using the Magnus series \cite{magnusseries}
\begin{equation}
    \frac{x}{e^x-1}=\sum_{k=0}^\infty \frac{B_k x^k}{k!}
\end{equation}
where $B_k$ are the Bernoulli numbers. The first three are $B_0=1$, $B_1=-1/2$ and $B_2=1/6$. The result is
\begin{equation}
   \frac{\partial}{\partial s_f} \chi(Q_+,Q_-,s_f,s_i)=\sum_{k=0}^{\infty}\frac{B_k}{k!}\{*,\chi\}^k\Big[-\mathcal{H}^I(Q_+,Q_-,s_f,s_i)\Big].
\end{equation}
If the interaction Hamiltonian is proportional to a small parameter $\epsilon$, the Magnusian can be expanded as
\begin{equation}\label{eq:perturbativeexpofmagnusian}
    \chi=\chi_{(1)}+\chi_{(2)}+\dots
\end{equation}
where $\chi_{(n)}\propto\epsilon^n$. The first-order Magnusian is
\begin{equation}
    \chi_{(1)}(Q_+,Q_-,s_f,s_i)=-\int_{s_i}^{s_f}ds \mathcal{H}^I(Q_+,Q_-,s,s_i).
\end{equation}
Using the expression for the interaction Hamiltonian (\ref{eq:interactionham}), we get
\begin{equation}\label{eq:magnusiandefinition}
    \chi_{(1)}(Q_+,Q_-,s_f,s_i)=-Q_-^B\int_{s_i}^{s_f}ds  F_A\big[X^{(0)}_{s,s_i}(Q_+)\big]M^A{}_B(s,s_i;Q_+).
\end{equation}
This is the main result of this paper. As explained before, in Appendix \ref{appendixnonlocal} we show that this formula continues to hold for nonlocal-in-time forces $F_A$, after the nonlocalities are treated perturbatively.

For completeness, we write below the second-order Magnusian
\begin{equation}
    \chi_{(2)}(Q_+,Q_-,s_f,s_i)=-\frac{1}{2}\int_{s_i}^{s_f}ds_1\int_{s_i}^{s_1}ds_2 \{\mathcal{H}^I(Q_+,Q_-,s_1,s_i),\mathcal{H}^I(Q_+,Q_-,s_2,s_i)\}.
\end{equation}

The interaction Hamiltonian in Eq. (\ref{eq:interactionham}) belongs to the class of linear generators introduced in Subsection \ref{sec:closure}. Because that class is closed under the doubled Poisson bracket, every nested bracket in the Magnus expansion remains linear in $Q_-$. Consequently, the Magnusian has the form 
\begin{equation}
\chi=Q_-^A\chi_A(Q_+,s_f,s_i)
\end{equation}
at every order. Also, its Poisson action on a physical observable $\O(Q_+)$ is independent of $Q_-$ so the exponentiated evolution is a function of $Q_+$ alone without requiring an additional physical limit.

\section{Application to leading 2.5PN radiation-reaction dynamics}\label{sec:2.5}

As an application, we consider the center-of-mass dynamics of a binary system at leading dissipative order in the post-Newtonian expansion. We work at leading order in the adiabatic expansion, in which the binary follows a Newtonian orbit on the orbital timescale while its orbital elements evolve slowly under the leading 2.5PN radiation-reaction force. Hence, the dynamics considered here are not the complete equations through 2.5PN order, but rather the leading dissipative correction to the Newtonian motion.

In dimensionless variables, the equations of motion are
\begin{subequations}\label{eq:2.5eoms}
\begin{eqnarray}
\dot{q}^i&=&p_i,\\
\dot{p}_i&=&-\frac{1}{q^2}n_i+F_i^{(2.5)}(\mathbf{q},\mathbf{p}).
\end{eqnarray}
\end{subequations}
Here, $q^i$ is the relative position, $q$ is its norm, $n_i=q^i/q$, and $p_i$ is the relative momentum. The dimensionless variables are related to their dimensionful counterparts by $q=\tilde{q}c^2/(GM)$, $p=\tilde{p}/(\mu c)$, and $t=\tilde{t}c^3/(GM)$, where $M$ and $\mu$ are the total and reduced masses, respectively.

In the harmonic gauge, the leading radiation-reaction force in center-of-mass variables is \cite{introPN}
\begin{eqnarray}
    F^{(2.5)}_i(\mathbf{q},\mathbf{p})&=&\frac{1}{ q^2}\left[\frac{24\nu}{5}\frac{1}{q}p_rp^2+\frac{136 \nu}{15}\frac{1}{q^2}p_r\right]n_i\nonumber\\
    &-&\frac{1}{ q^2}\left[\frac{8\nu }{5}\frac{1}{q}p^2+\frac{24\nu}{5} \frac{1}{q^2 }\right]p_i,
\end{eqnarray}
where we have used $p_r=\mathbf{n}\cdot \mathbf{p}$, $p^2=\mathbf{p}\cdot \mathbf{p}$, and the symmetric mass ratio $\nu=\mu/M$. Using the in-in formalism, the equations of motion in Eq.~\eqref{eq:2.5eoms} can be derived from the Hamiltonian
\begin{equation}
    \mathcal{H}(Q_+,Q_-)=\mathbf{p}_{+}\cdot\mathbf{p}_{-}+\frac{1}{q_+^2}\mathbf{n}_+\cdot\mathbf{q}_--q_-^i F^{2.5}_i(\mathbf{q}_+,\mathbf{p}_+).
\end{equation}

\subsection{Delaunay Variables}

To describe the zeroth-order dynamics, we use variables adapted to orbital motion. The radial motion can be parametrized as
\begin{equation}
    r=\frac{h}{1+e\cos f},
\end{equation}
where $h$ is the semilatus rectum, $e$ is the eccentricity and $f$ is the true anomaly. The radius can also be defined as
\begin{equation}
    r=a\big(1-e\cos u\big),
\end{equation}
where $a\equiv h/(1-e^2)$ is the semi-major axis and $u$ is the eccentric anomaly. The true and eccentric anomalies are related by 
\begin{subequations}\label{eq:relationfu}
\begin{eqnarray}
    \cos f &=& \frac{\cos u -e}{1-e\cos u},\\
    \sin f &=& \frac{\sqrt{1-e^2}\sin u}{1-e \cos u}.
\end{eqnarray}
\end{subequations}
The azimuthal angle is 
\begin{equation}
    \varphi= f+g,
\end{equation}
where $g$ is the argument of periastron. 
Following the notation in Ref.~\cite{Damour:2015isa}, we define the action variables
\begin{subequations}\label{eq:defactionvariableslg}
    \begin{eqnarray}
        \mathcal{L}&=&\sqrt{a},\\
        \mathcal{G}&=&\sqrt{a(1-e^2)}.
    \end{eqnarray}
\end{subequations}
In these variables, energy and angular momentum are
\begin{equation}\label{eq:defactionvariablesEJ}
    E=-\frac{1}{2\mathcal{L}^2}, \qquad J=\mathcal{G}.
\end{equation}
Lastly, we introduce the mean anomaly $l$, related to $u$ by Kepler's equation
\begin{equation}\label{eq:relationlu}
    l=u-e\sin u.
\end{equation}
The variables $(l,g,\dL,\dG)$ are called Delaunay variables and satisfy the canonical relation
\begin{equation}
    \{l,\dL\}=1,\quad \{g,\dG\}=1
\end{equation}
with all other brackets vanishing. 

Now, we perform a change of coordinates from $(r,\varphi,p_r,p_\varphi)$ to $(l,g,\mathcal{L},\mathcal{G})$. We will keep the dependence on $l$ implicit through $f(l,\dL,\dG)$, since the resulting expressions are easier to integrate over $f$ than over $l$. The coordinate transformation is
\begin{subequations}\label{eq:radialtodelaunay}
    \begin{eqnarray}
        r&=&\frac{\mathcal{G}^2}{1+e(\dL,\dG)\cos f(l,\dL,\dG)},\\
        \varphi&=& f(l,\dL,\dG)+g,\\
        p_r &=&\frac{e(\dL,\dG) \sin f(l,\dL,\dG)}{\mathcal{G}},\\
        p_\varphi&=& \mathcal{G},
    \end{eqnarray}
\end{subequations}
where the eccentricity is defined by $e=\sqrt{1-\mathcal{G}^2/\mathcal{L}^2}$. 

In Delaunay variables, the zeroth-order evolution is simply
\begin{subequations}\label{eq:delaunayflow}
    \begin{eqnarray}
        \dot{\dG}&=&0,\\
        \dot{\dL}&=&0,\\
        \dot{g} &=& 0,\\
        \dot{l}&=& \frac{1}{\dL^3}\label{eq:meananomalyhameq}.
    \end{eqnarray}
\end{subequations}
Thus, the Jacobi propagator $M^A{}_B$ in the $(l,g,\dL,\dG)$ basis is
\begin{equation}\label{eq:JcobiDelunay}
    M^A{}_B(t,t_i)=\begin{bmatrix}
        1 & 0 & -\frac{3}{\dL^4}(t-t_i) & 0\\0 &1 &0 &0\\0&0&1&0\\0&0&0&1
    \end{bmatrix}.
\end{equation}

\subsection{Interaction-picture Hamiltonian}

The term $q_-^iF_i^{(2.5)}(\mathbf{q}_+,\mathbf{p}_+)$ is a scalar under the change of phase-space coordinates. To linear order in the difference variables, it can therefore be written as
\begin{equation}
q_-^iF_i^{(2.5)}(\mathbf{q}_+,\mathbf{p}_+)=\mathcal{L}_-\mathcal{F}_{\mathcal{L}} +\mathcal{G}_-\mathcal{F}_{\mathcal{G}}
+l_-\mathcal{F}_l +g_-\mathcal{F}_g.
\end{equation}
The transformed force components are obtained by contracting the radial and angular components of the original force with the derivatives of $(r,\varphi)$ with respect to the Delaunay variables,
\begin{equation}\label{eq:forcesdelaunay}
    \mathcal{F}_A=\frac{\partial r}{\partial Q^A}F_r +\frac{\partial \varphi}{\partial Q^A}F_\varphi
\end{equation}
Here, the force components are $F_r=\mathbf{F}\cdot \mathbf{n}$ and $F_\varphi=r\mathbf{F}\cdot\hat{\mathbf{\varphi}}$. The coefficients relating spherical and Delaunay variables are obtained from Eq.~\eqref{eq:radialtodelaunay}. The expressions for $\mathcal{F}_A$ are straightforward but lengthy and will not be displayed.

\subsection{One-cycle Magnusian}

Inserting Eq. (\ref{eq:forcesdelaunay}) and Eq. (\ref{eq:JcobiDelunay}) into Eq. (\ref{eq:magnusiandefinition}), the Magnusian $\chi(Q_+,Q_-,s_f,s_i)$ is
\begin{eqnarray}
\chi(Q_+,Q_-,t_f,t_i)&=& \mathcal{L}_-\int_{t_i}^{t_f}\left[\mathcal{F}_{\mathcal{L}}-\frac{3(t-t_i)}{\mathcal{L}^4}\mathcal{F}_l\right]dt \\
&+&\mathcal{G}_-\int_{t_i}^{t_f}\mathcal{F}_{\mathcal{G}}dt
+l_-\int_{t_i}^{t_f}\mathcal{F}_ldt +g_-\int_{t_i}^{t_f}\mathcal{F}_gdt,\nonumber
\end{eqnarray}
where the integrands are evaluated on the zeroth-order flow (\ref{eq:delaunayflow}). Note that the first term has an extra contribution from $\mathcal{F}_l$ due to the Jacobi propagator. Also, the Magnusian is linear in the difference variables $Q_-$, which are retained during the Poisson bracket computation, although the final evolution of any physical observable is independent of them.

It will be convenient to choose $t_i$ and $t_f$ such that the true anomaly $f$ runs over one full cycle. This corresponds to the zeroth order period
\begin{equation}
T=2\pi \mathcal{L}^3,
\end{equation}
which is readily derived from Eq. (\ref{eq:meananomalyhameq}). Then, time integrals can be transformed using
\begin{equation}
dt =\frac{\mathcal{G}^3}{(1+e\cos f)^2}df
\end{equation}
so that they go from some initial $f_i(l,\mathcal{L},\mathcal{G})$ up to $f_i+2\pi$. The result is
\begin{equation}\label{eq:magnusianoneperiod}
    \chi_{\text{1-period}}(Q_+,Q_-)=l_-\chi_l+g_-\chi_g+\dL_-\chi_\dL+\dG_- \chi_\dG
\end{equation}
where each component is 
\begin{subequations}\label{eq:magnusiancomponents}
    \begin{eqnarray}
\chi_l&=&\dG^3\int_{f_i}^{f_i+2\pi}\mathcal{F}_l\frac{df}{\big(1+e\cos f\big)^2},\\
        \chi_g&=&\dG^3\int _{f_i}^{f_i+2\pi}\mathcal{F}_g  \frac{df}{\big(1+e\cos f\big)^2},\\
        \chi_\dL&=&\dG^3\int _{f_i}^{f_i+2\pi}\left[\mathcal{F}_\dL-\frac{3\big(l^{(0)}(f)-l\big)}{\dL}\mathcal{F}_l\right]  \frac{df}{\big(1+e\cos f\big)^2},\\
        \chi_\dG&=&\dG^3\int _{f_i}^{f_i+2\pi} \mathcal{F}_\dG  \frac{df}{\big(1+e\cos f\big)^2}.
    \end{eqnarray}
\end{subequations}
Note that the first-order Magnusian is evaluated on the zeroth-order flow and so we replaced the term $t-t_i$ by $\mathcal{L}^3\big[l^{(0)}(f)-l\big]$. The term $l$ is a phase-space variable of the Magnusian while $l^{(0)}(f)$ is to be integrated.

The terms proportional to $\mathcal{F}_\mathcal{L}$ and $\mathcal{F}_\mathcal{G}$ are periodic in $f$ so the limits of integration can be shifted by $f_i$. Then, both vanish since they are odd in $f$. The term proportional to $l^{(0)}(f)$ can be integrated using that the mean anomaly has a reflection symmetry
\begin{equation}
    l^{(0)}(2\pi-f)=2\pi-l^{(0)}(f).
\end{equation}
The integrals in Eq. (\ref{eq:magnusiancomponents}) evaluate to
\begin{subequations}\label{eq:Magnusiancomponentssolved}
    \begin{eqnarray}
        \chi_l&=&-\frac{2\pi \nu \dL^3}{15\dG^7}P_0(e),\\
         \chi_\dL&=&\frac{2\pi \nu \dL^2}{5\dG^7}\Big\{P_0(e)\big[\pi+f(l,\mathcal{L},\mathcal{G})-l\big]+\sum_{n=1}^4P_n(e)\frac{\sin(nf)}{n}{}\Big\},\\
        \chi_g &=&-\frac{8\pi \nu }{5\dG^4}P_5(e),\\
   \chi_\dG&=&0.
    \end{eqnarray}
\end{subequations}
where the polynomials $P_n$ are
\begin{subequations}
\begin{eqnarray}
P_0(e)&=&96+292e^2+37e^4,\\
P_1(e)&=&408e+346e^3+6e^5,\\
P_2(e)&=&380e^2+120e^4,\\
P_3(e)&=&182e^3+18e^5,\\
P_4(e)&=&35e^4\\
P_5(e)&=&8+7e^2.
\end{eqnarray}
\end{subequations}
Let us emphasize that the components $\chi_A$ of the first-order Magnusian $\chi_{(1)}$ in Eq. (\ref{eq:Magnusiancomponentssolved}) are functions of the Delaunay phase-space variables $(l,\mathcal{L},g,\mathcal{G})$. The eccentricity $e(\mathcal{L},\mathcal{G})$ and the true anomaly $f(l,\mathcal{L},\mathcal{G})$ only appear as functions of the other variables.

Once the Magnusian has been determined, the one-cycle evolution of the system can be constructed as follows. Let $Q_0$ be the initial phase-space state. We first act on $Q_0$ with the Magnusian map to produce its interaction picture representation
\begin{equation}
    Q_0^I=e^{\{\chi,*\}}Q_0
\end{equation}
At this stage no physical time evolution has yet taken place: $Q_0^I$ is the interaction-picture representation of the initial state. As discussed in Section \ref{sec:interaction}, it should be interpreted as the initial state whose unperturbed evolution reaches the same endpoint that $Q_0$ would reach under the full dynamics. The second step is therefore to evolve $Q^I_0$ with the zeroth-order flow for one orbital period $T_0=2\pi \mathcal{L}_0^3$. This flow acts trivially on $\mathcal{L}$, $\mathcal{G}$ and $g$, but advances the mean anomaly according to
\begin{equation}
    l^I_0 \rightarrow l^I_0 +\frac{T_0}{\big(\mathcal{L}^I_0\big)^3}.
\end{equation}
Crucially, the result differs from the initial mean anomaly $l_0$ by two terms. First, the second term is in general not exactly equal to $2\pi$. Second, $l_0^I=l_0-\chi_L+\dots$. The endpoint generated by the Magnusian map followed by the zeroth-order flow therefore does not, in general, lie at precisely the same orbital phase as the initial point. Athough the mismatch is perturbatively small, it would accumulate over several cycles.

To construct a true one-cycle map between points of equal orbital phase, it is natural to modify the zeroth-order evolution and add a small time interval $\Delta T$, chosen so that the orbital phase returns to its initial value modulo $2\pi$. Since $\Delta T=O(\epsilon)$, replacing the time interval entering the Magnusian by the phase-matched interval $T_0+\Delta T$ would modify the Magnusian only at $O(\epsilon^2)$. Consequently, at first perturbative order the Magnusian may consistently be evaluated over the original interval $T$, while the modified zeroth-order evolution is used to assign the endpoint to the desired orbital phase.

The complete evolution is obtained by iterating this construction. Thus, when the initial data are specified at periastron, the resulting discrete map relates the system at successive periastron passages. More generally, choosing any other initial orbital phase defines an analogous map between corresponding points on successive orbital cycles.

As a consistency check, the Magnusian coefficients reproduce the standard orbit-averaged evolution induced by leading-order gravitational radiation reaction. At first order in the Magnusian map, the changes in the Delaunay actions are simply
\begin{equation}
    \Delta \mathcal{L} = \chi_l,
    \qquad
    \Delta \mathcal{G} = \chi_g.
\end{equation}
Using Eq.  (\ref{eq:defactionvariablesEJ}), we can calculate the change in energy and angular momentum in terms of the rates of change of action variables as $\Delta E = \chi_l/\mathcal{L}^3$ and $\Delta J = \chi_g$. Then, using that the period is  $T = 2\pi \mathcal{L}^3$, we obtain
\begin{equation}
    \left\langle \frac{dE}{dt} \right\rangle
    =-\frac{32}{5}\frac{\nu}{a^5 (1-e^2)^{7/2}}\Big[1+\frac{73}{24}e^2+\frac{37}{96}e^4\Big]
\end{equation}
and
\begin{equation}
    \left\langle \frac{dJ}{dt} \right\rangle
    =-\frac{32}{5}\frac{\nu}{a^{7/2}(1-e^2)^2}\Big[1+\frac{7}{8}e^2\Big]
\end{equation}
which are precisely the standard Peters fluxes \cite{Peters:1963ux,Peters:1964zz}  of an eccentric binary at quadrupole order\footnote{We thank Lidia Gomes Da Silva for pointing this out.}. The remaining Magnusian coefficients have a similarly direct interpretation: $\chi_{\mathcal{G}}=0$ implies $\Delta g=0$, as expected in the absence of conservative periastron precession at the order considered here, while $\chi_{\mathcal{L}}$ generates a shift in the mean anomaly. Thus, the one-cycle Magnusian packages the familiar secular losses of energy and angular momentum, together with the associated phase evolution, into a single phase-space generator.

\begin{figure}[h!]
\centering
\begin{subfigure}[b]{0.49\textwidth}
\centering
\hspace{-0.08\textwidth}
\includegraphics[width=\linewidth]{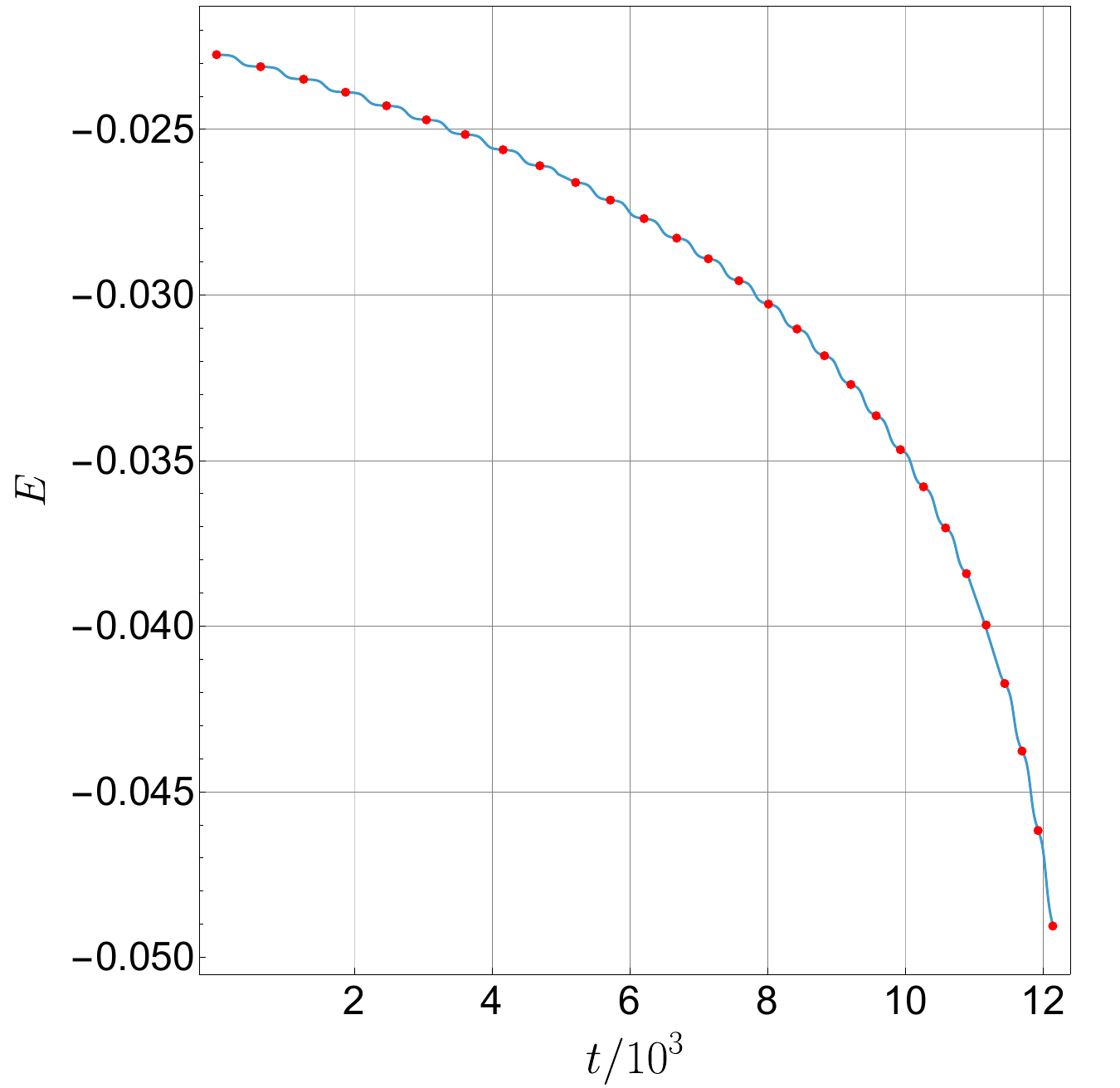}

\end{subfigure}
\hfill
\begin{subfigure}[b]{0.49\textwidth}
\centering
\includegraphics[width=\linewidth]{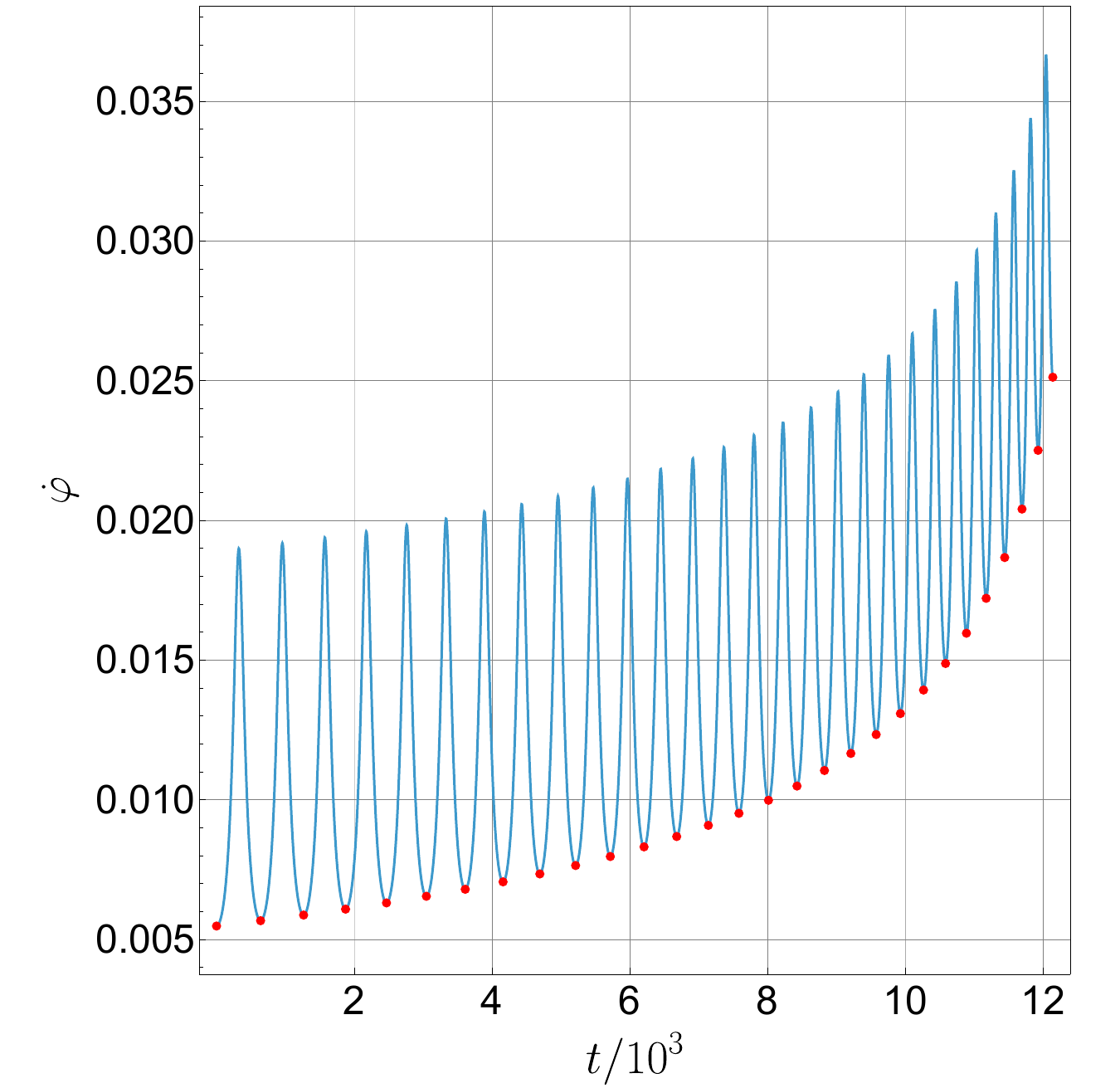}

\end{subfigure}
\caption{Comparison between the numerical and Magnusian evolutions of the energy \textbf{(left)} and angular frequency \textbf{(right)} with initial true anomaly $f_0=\pi$. The Magnusian evolution is truncated at four Poisson brackets.}
\label{fig:time-evolution}
\end{figure}

As an example, we choose a binary with symmetric mass ratio $\nu=1/4$ and initial conditions $e=0.3$, $h=20$, and $f=\pi$ and $g=0$. In Fig.~\ref{fig:time-evolution}, the Magnusian evolution of the energy $E$ and angular frequency $\dot{\varphi}$, shown in red, is compared with a numerical solution of the Newtonian equations supplemented by the leading 2.5PN radiation-reaction force, shown in blue.

\begin{figure}[h!]
    \centering
   \includegraphics[width=0.49\textwidth]{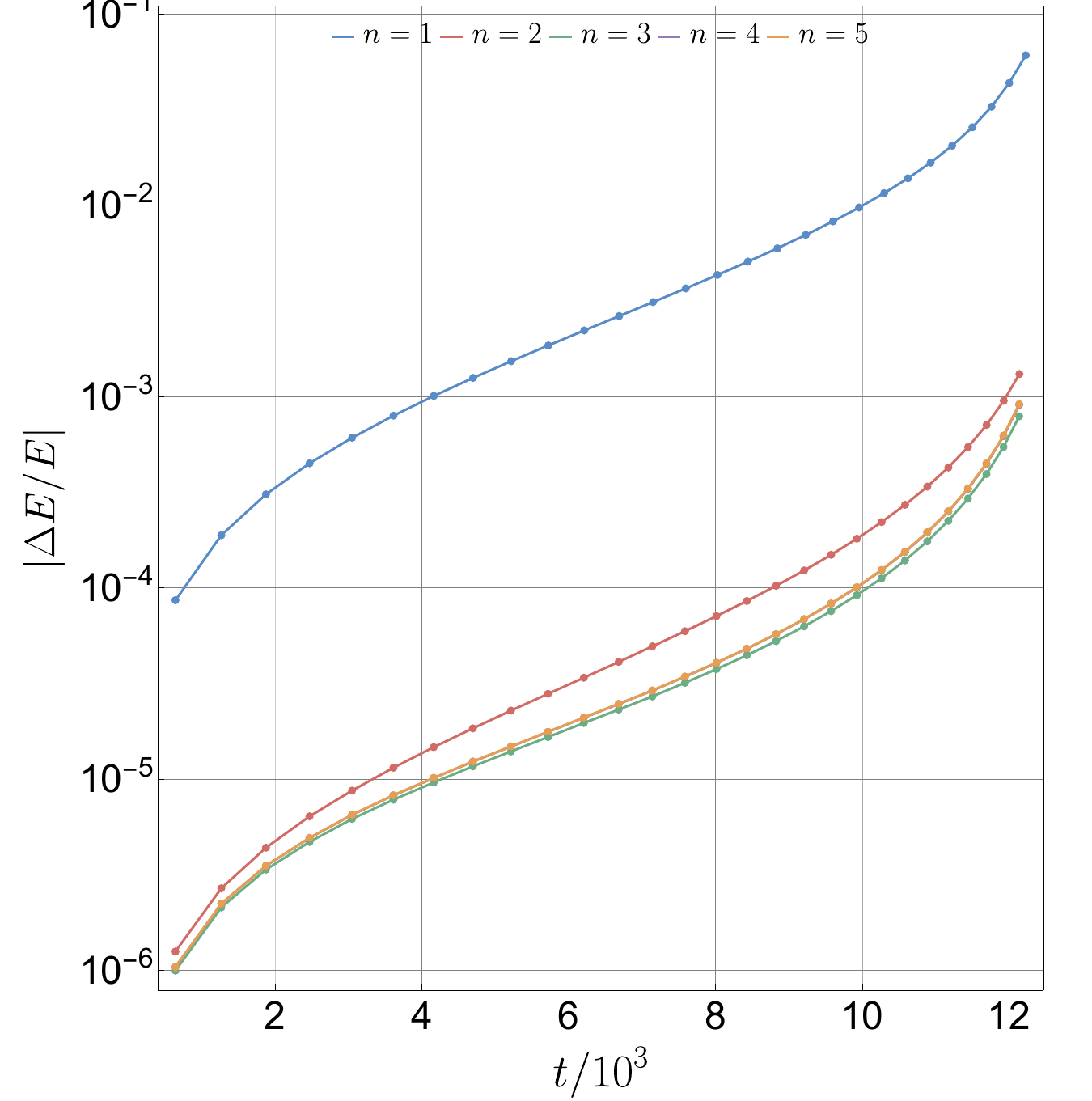}
    \hfill
     \includegraphics[width=0.49\textwidth]{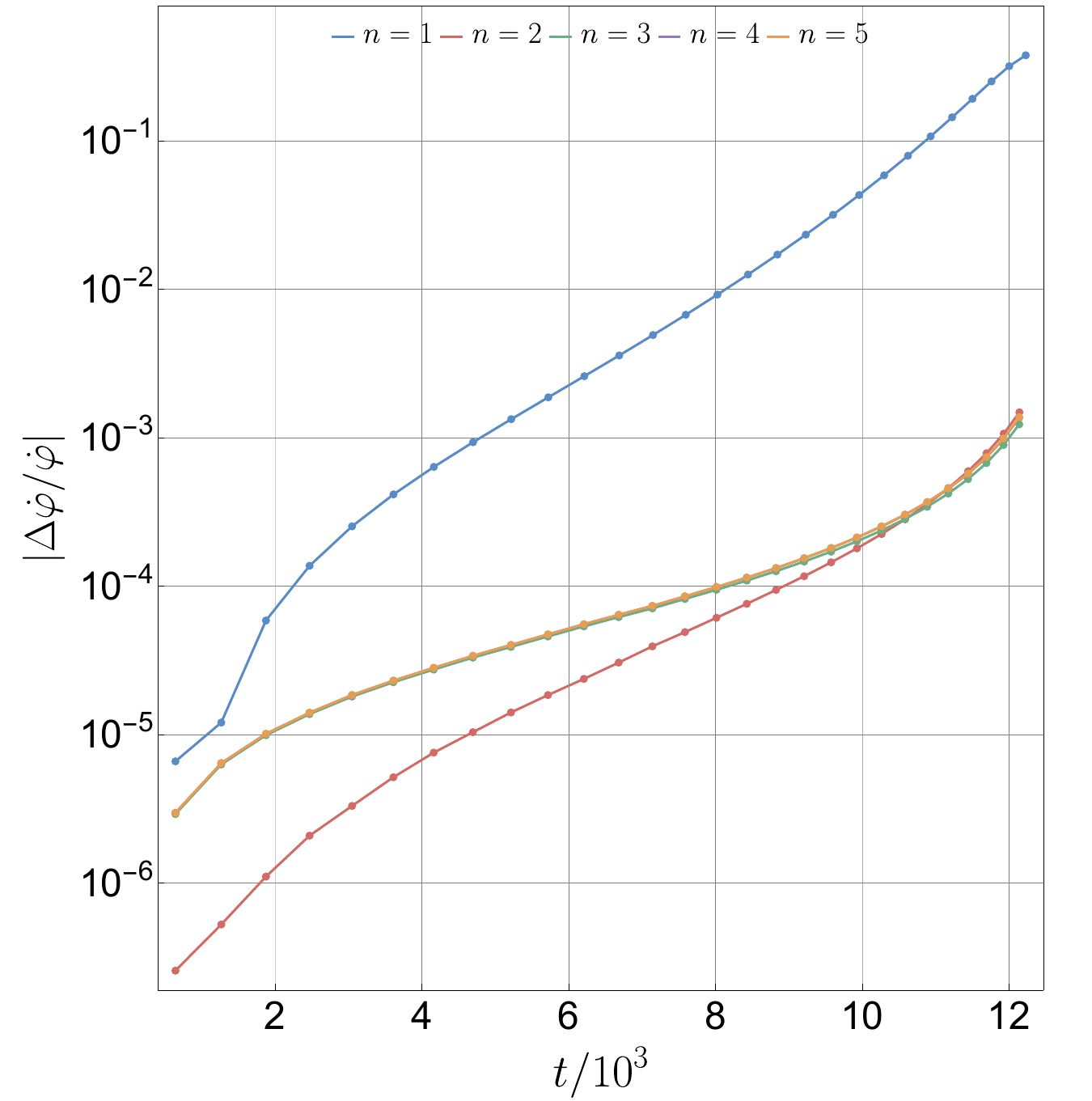}

    \caption{Relative errors between the Magnusian and numerical evolutions of energy \textbf{(left)} and angular frequency \textbf{(right)}, for different numbers $n$ of nested Poisson brackets.}
    \label{fig:errors}
\end{figure}

The finite-time evolution is calculated using Eq.~(\ref{eq:magnusianevolutioninteraction}), where we exponentiate the first-order Magnusian and truncate the resulting series after four nested Poisson brackets. This procedure partially resums the finite transformation generated by $\chi_{(1)}$, while neglecting the independent contributions from $\chi_{(2)}$ and higher-order terms in the Magnus expansion. Figure~\ref{fig:errors} illustrates the convergence of this truncation by comparing the absolute vaue of the relative errors in the energy and angular frequency for different numbers $n$ of nested Poisson brackets. The error decreases rapidly as $n$ is increased, but the improvement saturates at approximately $n=4$, with additional brackets producing negligible improvement. This indicates that beyond this point the dominant error is no longer associated with truncating the exponential generated by $\chi_{(1)}$, but rather with the absence of $\chi_{(2)}$. The fact that lower truncation orders, such as $n=2$ or $n=3$, can occasionally yield slightly smaller errors should therefore not be interpreted as better convergence. These truncations may fortuitously overshoot the $\chi_{(1)}$ contribution and partially compensate for the neglected higher-order terms.

As explained above, the Magnusian evolution in Fig. \ref{fig:time-evolution} follows the apoastron passage, at $f=\pi$. Choosing initial conditions with true anomaly $f=0$ instead yields an periastron-to-periastron map. More generally, one could numerically solve only the first orbit, sample a set of points along it, and subsequently evolve each of these points with the Magnusian map. In this way, the evolution can be reconstructed at arbitrary orbital phases without numerically integrating each subsequent orbit.

\section{Effective Hamiltonian interpretation and relation to near-identity transformations}\label{sec:nit}

The Magnusian admits two complementary interpretations. As discussed in the preceding sections, it generates the finite canonical transformation that evolves observables from an initial time $s_i$ to a final time $s_f$. However, it can also be used to define an effective time-independent Hamiltonian whose evolution over the same interval reproduces this finite transformation. This second interpretation establishes a direct connection\footnote{We thank Jan Steinhoff for clarifying this connection.} between the Magnusian and near-identity transformations (NITs).

A NIT is a perturbative redefinition of the dynamical variables designed to absorb the oscillatory part of the motion, leaving transformed variables that evolve secularly. For a Hamiltonian system with a periodic perturbation, this redefinition may be implemented as a time-dependent canonical transformation that removes the explicit time dependence of the Hamiltonian order by order. The resulting time-independent Hamiltonian governs the secular evolution from one period to the next. As we show below, this effective Hamiltonian is proportional to the 1-period Magnusian. The Magnusian therefore provides a canonical realization of the averaged dynamics obtained through a NIT.

This construction should be distinguished from the NITs commonly employed in the self-force literature. Those transformations are formulated directly at the level of the equations of motion, which need not be Hamiltonian, and are typically chosen to eliminate dependence on all fast variables, usually the orbital angles in an action-angle formulation. By contrast, the canonical construction considered here eliminates the explicit dependence of the Hamiltonian on the time variable.

\subsection{Overview of NITs}

As a simple example, we consider  the one-dimensional differential equation
\begin{equation}\label{eq:NIT1D}
    \frac{dx}{dt}=\epsilon f(x,t)
\end{equation}
where $\epsilon$ is a small parameter and $f(x,t)$ is periodic in $t$ with period $T$. We decompose the forcing term into its average and oscillatory parts
\begin{equation}
    f(x,t)=\langle f\rangle_T +f_\text{osc}(x,t)
\end{equation}
where the average is defined by
\begin{equation}
    \langle f\rangle_T=\frac{1}{T}\int_0^T dt' f(x,t').
\end{equation}
By construction, the oscillatory contribution has vanishing average,
\begin{equation}
\int_0^T dt  f_{\mathrm{osc}}(x,t)=0.
\end{equation}
It is possible to remove the oscillatory behaviour of $f(x,t)$ order by order in $\epsilon$ by doing a time-dependent coordinate transformation
\begin{equation}
    y=x-\epsilon u(x,t)+O(\epsilon^2).
\end{equation}
In the new coordinates, Eq. (\ref{eq:NIT1D}) becomes
\begin{equation}
    \frac{dy}{dt}=\epsilon\big[ f-\frac{\partial u}{\partial t}\big] +O(\epsilon^2)
\end{equation}
where the right-hand side is understood to be evaluated on $x(y,t)$. To eliminate $f_\text{osc}$, we choose
\begin{equation}\label{eq:NITfunction}
    u(x,t)=\int_0^{t}f_{\text{osc}}(x,t')dt'
\end{equation}
so that the transformed equation is
\begin{equation}
    \frac{dy}{dt}=\epsilon \langle f\rangle_T+O(\epsilon^2).
\end{equation}
The transformed variable therefore evolves according to the averaged force. Moreover, because $f_{\mathrm{osc}}$ is periodic and has vanishing average, the function $u$ is also periodic and satisfies
\begin{equation}
    u(x,0)=u(x,T)=0.
\end{equation}
Hence, the original and transformed variables agree at the beginning and end of each cycle, although they differ by an oscillatory correction within the cycle. This procedure can be extended order by order in $\epsilon$, with the NIT at each order chosen to remove the oscillatory terms generated at that order.

The same principle can be implemented canonically when the original dynamics are Hamiltonian. In that case, the NIT removes the explicit time dependence of the Hamiltonian, and the resulting time-independent Hamiltonian can be identified directly with the Magnusian. We discuss this in the next subsection.

\subsection{Canonical NITs and the Magnusian as an effective Hamiltonian}

We now review the application of NITs to Hamiltonian systems. Consider a time-dependent Hamiltonian of the form 
\begin{equation}
    H(q,p,s)=\epsilon V(q,p,s)
\end{equation}
where the perturbation $V$ is periodic in $s$ with period $T$. Let $G(q,p,s)=O(\epsilon)$ generate a time-dependent canonical transformation. To first order in $\epsilon$, the transformed Hamiltonian is
\begin{equation}
H'(q',p',s)=H(q,p,s)-\frac{\partial G(q,p,s)}{\partial s}
+O(\epsilon^2).
\end{equation}
In a manner analogous to Eq. (\ref{eq:NITfunction}), we choose the generator to be
\begin{equation}G(q,p,s)=\epsilon\int_0^{s}\left[V(q,p,s')-\langle V\rangle_T \right]ds'
\end{equation}
Then the new Hamiltonian is
\begin{equation}
    H'=\epsilon \langle V \rangle _T+O(\epsilon^2).
\end{equation}
The time-dependent canonical transformation thus absorbs the oscillatory dynamics within each period, while the transformed, time-independent Hamiltonian governs the accumulated evolution from one period to the next. Also, since $G(q,p,s)$ vanishes on $s=0$ and $s=T$, the new coordinates match the old ones at the end of each cycle.

Beyond first order, the generating function $G(q,p,s)$ must be constructed iteratively to eliminate the time-dependence of the Hamiltonian, order by order in $\epsilon$. We do not carry out this nonperturbative canonical transformation explicitly. Instead, we show that the resulting Hamiltonian must be proportional to the Magnusian (See Eq. (\ref{eq:NITmagnusian}) below).

First, let $H'(q,p;T)$ denote the time-independent Hamiltonian obtained after carrying the canonical NIT to all orders. It depends parametrically on the period $T$, but it is independent of time. Recall that the time-dependent canonical transformation is chosen to reduce to the identity at the beginning and end of each period. Consequently, the evolution generated by $H'$ over one period agrees with that generated by the original time-dependent Hamiltonian. Furthermore, it is time-independent, and hence its action on phase-space observables can be exponentiated to be
\begin{equation}
    \O_f=e^{-T\{H',*\}}\O_i
\end{equation}
On the other hand, the defining evolution generated by the 1-period Magnusian is
\begin{equation}
    \O_f=e^{\{\chi,*\}}\O_i
\end{equation}
Choosing the perturbative branch continuously connected to the identity, the generators may therefore be identified as
\begin{equation}\label{eq:NITmagnusian}
    H'=-\frac{\chi}{T}.
\end{equation}
Thus, although we do not construct the nonperturbative canonical NIT explicitly\footnote{The construction of this canonical transformation to all orders will be presented elsewhere.}, its resulting time-independent Hamiltonian is fixed by the requirement that it reproduce the same one-period evolution as the original system. It is therefore precisely the effective Hamiltonian defined by the Magnusian in Eq. (\ref{eq:NITmagnusian}).

\subsection{Comparison with the use of NITs in self-force theory}

We now compare the canonical construction described above with the NITs commonly employed in self-force calculations. Following Ref.~\cite{van_de_Meent_2018,Lynch_2022,lynch2022efficient}, consider a system whose zeroth-order dynamics are integrable and may therefore be described in terms of action-angle variables $(\theta^i,I_i)$. In the presence of a perturbation, the equations of motion take the form
\begin{subequations}
\begin{eqnarray}
\frac{d\theta^i}{d\lambda} &=& \Upsilon^i(I) + \epsilon f_{(1)}^i(\theta,I) +
O(\epsilon^2), \\
\frac{dI_i}{d\lambda}
&=& \epsilon F^{(1)}_i(\theta,I)+O(\epsilon^2),
\end{eqnarray}
\end{subequations}
where $\Upsilon^i(I)$ are the frequencies of the unperturbed motion, and $f_{(1)}^i$ and $F_i^{(1)}$ describe the leading perturbative forcing.

The oscillatory piece of the forcing terms, given by their dependence on the angle variables $\theta^i$, can be absorbed by a NIT. We define new variables $\tilde{\theta}^i$ and $\tilde{I}_i$ by
\begin{subequations}
\begin{eqnarray}
    \theta^i&=&\tilde{\theta}^i-\epsilon X^i(\tilde{\theta},\tilde{I})+O(\epsilon^2),\\
    I_i&=&\tilde{I}_i-\epsilon Y_i (\tilde{\theta},\tilde{I})+O(\epsilon^2).
\end{eqnarray}
\end{subequations}
where the functions $X^i$ and $Y_i$ can be chosen, order by order, to remove the oscillating pieces of the forcing terms $f^i$ and $F_i$. 

To determine these functions, the forcing terms are expanded in Fourier modes
\begin{subequations}
    \begin{eqnarray}
        f_{(1)}^i(\theta,I)&=&\sum_{\vec{k}}f^i_{(1),\vec{k}}(I)e^{i \vec{k}\cdot \vec{\theta}}\\
        F_i^{(1)}(\theta,I)&=&\sum_{\vec{k}}F_{i,\vec{k}}^{(1)}(I)e^{i \vec{k}\cdot \vec{\theta}}
    \end{eqnarray}
\end{subequations}
We can eliminate each Fourier mode with $\vec{k}\neq 0$ by choosing the Fourier modes of the functions $X^i$ and $Y_i$ to be
\begin{subequations}\label{eq:generatorsnits}
    \begin{eqnarray}
        X^i_{\vec{k}}(I) &=&\frac{i}{\vec{k}\cdot \vec{\Upsilon}}f_{(1)\vec{k}}^i+\frac{1}{\big(\vec{k}\cdot\vec{\Upsilon}\big)^2}\frac{\partial \Upsilon^i}{\partial I_j}F^{(1)}_{j,\vec{k}},\\
        Y_{i,\vec{k}}(I) &=&\frac{i}{\vec{k}\cdot \vec{\Upsilon}}F^{(1)}_{i,\vec{k}}.
    \end{eqnarray}
\end{subequations}
The equations of motion become
\begin{subequations}
    \begin{eqnarray}
             \frac{d\tilde{\theta}^i}{d\lambda}&=&\Upsilon^i(\tilde{I})+\epsilon \langle f_{(1)}^i\rangle(\tilde{I})+O(\epsilon^2) \\
        \frac{d\tilde{I}_i}{d\lambda}&=&\epsilon \langle F^{(1)}_i\rangle (\tilde{I})+O(\epsilon^2)
    \end{eqnarray}
\end{subequations}
where we defined the average over all angle variables 
\begin{equation}
    \langle F \rangle=\frac{1}{(2\pi)^N}\int d^N\theta F(\theta,I).
\end{equation}
The transformed variables therefore satisfy equations that are independent of the fast angles and describe the averaged evolution on the slow phase space.

This construction and the Magnusian formalism share the same broad objective: both reorganize the dynamics to separate rapidly oscillating motion from the evolution accumulated over longer timescales. They differ, however, in the variables that are eliminated and in the form of the resulting dynamics. The NIT used above removes dependence on all fast angles and produces continuous averaged equations of motion for the transformed variables. It is formulated directly at the level of the equations of motion and does not require the original system to possess a Hamiltonian structure. By contrast, the canonical construction associated with the Magnusian removes the explicit dependence of the effective Hamiltonian on the evolution parameter over a chosen interval. In a multiperiodic system, choosing an interval corresponding to one particular orbital cycle does not, in general, average uniformly over the full angle torus. The phases of the remaining angles therefore continue to appear in the finite-time map and evolve from one cycle to the next.

This distinction has important consequences near resonances. The Fourier modes in Eq.~(\ref{eq:generatorsnits}) contain factors of $\vec{k}\cdot\vec{\Upsilon}$ in their denominators. When
\begin{equation}
\vec{k}\cdot\vec{\Upsilon}=0
\end{equation}
for some nonzero $\vec{k}$, the corresponding NIT becomes singular and the resonant modes must be retained and treated separately. The finite-time integrals defining the Magnusian do not face this problem and therefore remain finite at resonances.

Lastly, we emphasize that the finite-time-map formulation of the Magnusian may also offer computational advantages. Once the Magnusian has been determined, observables can be propagated from one cycle to the next by exponentiating the action of its Poisson bracket, without reintegrating the original differential equations over every period. Moreover, the Magnus series organizes the finite-time evolution systematically in perturbation theory: higher-order corrections are generated by nested Poisson brackets, which might offer an advantage over having to compute the corrections $X^i$ and $Y_i$ to higher and higher orders.

\section{Discussion and outlook}\label{sec:disc}

In Section \ref{sec:2.5}, we focused on the Magnusian associated with the secular evolution of a bound inspiral. The resulting generator defines a map on phase space, relating the orbital variables at the beginning and end of each radial cycle. This construction is not restricted to periodic dynamics, however. For an unbound orbit, one may instead define a scattering Magnusian by integrating along the full trajectory from the asymptotic past to the asymptotic future. It would be interesting to understand more systematically how these two objects are related. In particular, one may ask whether the bound and scattering Magnusians can be obtained as different analytic continuations of a common phase-space function, and whether the evolution of bound observables can be reconstructed from scattering data, or vice versa. Such a relation would provide an interesting link to the connections between bound and unbound dynamics that have appeared in several formulations of the relativistic two-body problem \cite{K_lin_2020,K_lin_2020II,Cho_2022}. 

A second natural direction is to extend the present calculation to higher orders in the post-Newtonian expansion. Since the in-in formulation developed here accommodates nonlocal forces directly (as explained in Appendix \ref{appendixnonlocal}), the 4PN dynamics can in principle be incorporated into the Magnusian.

The same framework may also be extended beyond the post-Newtonian approximation to the gravitational self-force problem. After integrating out the gravitational field, the self-force dynamics is naturally expressed in terms of nonlocal worldline interactions involving retarded $n$-point functions. A central difficulty is that these $n$-point functions are singular when evaluated on the worldline and must be regularized before they can be used to construct a finite effective force. The regularization of the relevant $n$-point functions developed in Ref.~\cite{Harte:2025tmd} may provide the necessary input for defining a regularized self-force Magnusian. This would allow the conservative and dissipative self-force dynamics to be encoded in a phase-space generator and could provide a systematic method for constructing finite-time or one-period maps beyond leading order in the mass ratio.

Lastly, it is necessary to compare the computational cost and long-time accuracy of iterated Magnusian maps with existing NIT-based inspiral schemes \cite{van_de_Meent_2018,Lynch_2022,lynch2022efficient}, particularly near orbital resonances. 

\section*{Acknowledgments}

We thank Raj Patil, Trevor Scheopner and Jan Steinhoff for discussions about the Magnusian formalism. We also thank Aldo Gamboa for providing the numerical code to solve the 2.5PN dynamics. We thank Lidia Gomes Da Silva for many fruitful discussions on the numerical implementation of Magnusian maps.
\emph{Funded by the European Union. Views and opinions expressed are however those of the author(s) only and do not necessarily reflect those of the European Union or the European Research Council Executive Agency. Neither the European Union nor the granting authority can be held responsible for them. This work is supported by ERC grant (GWSky/ 101167314).}


\bibliographystyle{JHEP}
\bibliography{Ref}

\appendix

\section{Integrating out a scalar field: A Self-Force Example}\label{appendixscalar}

In this appendix, we illustrate the procedure described in Subsection \ref{sec:mediatingfield} by explicitly constructing the effective action obtained after integrating out a scalar field. The resulting effective dynamics are generically nonlocal in time. In Appendix \ref{appendixnonlocal}, we show how these nonlocal contributions can be order reduced, perturbatively, to obtain an equivalent local Hamiltonian description.

We consider a point particle of mass $m_0$ and scalar charge $k$ moving on an arbitrary curved background $g_{\mu\nu}$ under the influence of a scalar field $\phi$. We use the Lagrangian formulation of the dynamics for simplicity, but the Hamiltonian formulation achieves the same result.

The action functional of the system is
\begin{equation}
S[q,\phi]=S_q[q]+S_\phi[\phi]+S_{\text{int}}[q,\phi]
\end{equation}
where
\begin{subequations}
\begin{eqnarray}
S_q &=& -m_0\int d\tau,\\
S_\phi &=& \int \left\{\frac{1}{2}g^{\mu\nu}\nabla_\mu \phi \nabla_\nu \phi +U(\phi)\right\}dV,\\
S_{\text{int}}&=&\int\rho(x) \phi(x) dV.
\end{eqnarray}    
\end{subequations}
Here, $dV=\sqrt{-g}d^4x$ and
\begin{equation}
    \rho(x)=k\int \delta^{(4)}(x,q_\tau)d\tau
\end{equation}
is the scalar charge density of the point particle.

Varying the action yields the equations of motion
\begin{subequations}
    \begin{eqnarray}
        D_\tau \big[mu_\mu\big]&=&k\nabla_\mu \phi, \\
        \Box \phi - U'(\phi)&=& \rho
    \end{eqnarray}
\end{subequations}
where we have defined the effective mass $m=m_0-k\phi(q)$ and $u^\mu\equiv dq^\mu/d\tau$.

To integrate out the scalar field, we follow the formalism explained in Section~\ref{sec:galley}. We start by doubling the dynamical degrees of freedom, so that the action is a functional of $q_1$, $q_2$, $\phi_1$ and $\phi_2$
\begin{equation}
    S_{\text{double}}[q_a,\phi_a]=S_q[q_1]-S_q[q_2]+S_\phi[\phi_1]-S_\phi[\phi_2]+S_\text{int}[q_1,\phi_1]-S_\text{int}[q_2,\phi_2].
\end{equation}
The next step is to use $\pm$ variables $q_\pm$ and $\phi_\pm$ and expand the action to linear order in the difference variables. Let us focus on the terms involving the scalar field, which we want to integrate out.

The scalar field action functional is
\begin{equation}
    S_\phi[\phi_1]-S_{\phi}[\phi_2]=\int \left\{-\phi_-\Box \phi_+ +U'(\phi_+)\phi_-\right\}dV+O(\phi_-^2).
\end{equation}
The interaction term becomes
\begin{equation}
    S_\text{int}[q_1,\phi_1]-S_\text{int}[q_2,\phi_2]=\int \big\{\rho_- \phi_+ +\rho_+ \phi_-\big\}dV.
\end{equation}
Here, $\rho_\pm$ are defined analogously to $q_\pm$. Importantly, $\rho_+$ is \emph{not} the charge density of $q_+$. It is the half-sum of the charge densities of $q_1$ and $q_2$. 

Our final action principle is then
\begin{equation}
    S_\text{double}[q_a,\phi_a]=S_q[q_1]-S_q[q_2]+\int \left\{-\phi_-\Box \phi_+ +U'(\phi_+)\phi_-\right\}dV+\int \big\{\rho_- \phi_+ +\rho_+ \phi_-\big\}dV
\end{equation}

In the $\pm$ variables, the field equations are
\begin{subequations}
    \begin{eqnarray}
        \Box \phi_+ - U'(\phi_+)&=&\rho_+,\label{eq:solution+field}\\
        \Big[\Box - U''(\phi_+)\Big]\phi_-&=&\rho_-.
    \end{eqnarray}
\end{subequations}
As explained around Eqs.~\eqref{eq:variation-minus} and \eqref{eq:variation-plus}, $\phi_+$ satisfies the same field equations as the physical field, while $\phi_-$ solves the field equations for linearized perturbations.

Substituting the field equation for $\phi_+$ back into the action, we see that the terms proportional to the field equation cancel, leaving the term $\rho_-\phi_+$, where $\phi_+$ denotes the solution of Eq.~\eqref{eq:solution+field}. The effective action for the matter degrees of freedom is then
\begin{equation}
    S_\text{eff}[q_a]=S_q[q_1]-S_q[q_2]+\int \rho_- \phi_+[\rho_+] dV.
\end{equation}
Next, we solve the field equation for $\phi_+$ perturbatively by expanding
\begin{equation}
    \phi_+ = \bar{\phi}+\phi_+^{(1)}+\phi_+^{(2)}+\dots
\end{equation}
The successive orders then satisfy
\begin{subequations}
    \begin{eqnarray}
        \Box \bar{\phi}- U'(\bar{\phi})&=&0,\\
        D\phi_+^{(1)}&=&\rho_+,\\
        D\phi_+^{(2)}&=&\frac{\bar{U}'''}{2}\big[\phi_+^{(1)}\big]^2.
    \end{eqnarray}
\end{subequations}
Here, $D\equiv\Box-\bar{U}''$ and $\bar{U}\equiv U(\bar{\phi})$.

We solve the field equations for $\phi_+^{(n)}$ iteratively using the retarded Green's function
\begin{equation}
    D G_\text{ret}(x,y)=\delta^{(4)}(x,y).
\end{equation}
The choice of causality is not arbitrary: In the doubled phase-space formalism, the average field satisfies initial boundary conditions so it must be integrated using the retarded Green's function \cite{Galley1,Galley2}.

This gives
\begin{subequations}
    \begin{eqnarray}
        \phi_+^{(1)}(x)&=&\int dV_y G_\text{ret} (x,y) \rho_+ (y),\\
        \phi_+^{(2)}(x)&=&\int dV_ydV_z J_\text{ret}(x,y,z) \rho_+(y) \rho_+(z)
    \end{eqnarray}
    \end{subequations}
where we defined the retarded 3-point function
\begin{equation}
    J_\text{ret}(x,y,z)=\int dV_w \frac{\bar{U}'''(w)}{2}G_\text{ret}(x,w)G_\text{ret}(w,y)G_\text{ret}(w,z).
\end{equation}
The resulting on-shell doubled action is
\begin{eqnarray}\label{eq:effectiveactionscalar}
    S_\text{double/on-shell}&=&S_q[q_1]-S_q[q_2]+\int dV \rho_- \bar{\phi}\nonumber\\
    &+& \int dV_xdV_y \rho_- (x) G_\text{ret} (x,y) \rho_+(y) \\\nonumber
    &+& \int dV_x dV_y dV_z \rho_-(x) J_\text{ret}(x,y,z)\rho_+(y) \rho_+(z)
\end{eqnarray}
This construction can be continued straightforwardly to arbitrarily high orders in $\rho_+$.

The expression above is understood as a bare effective action. For a point particle, the coincidence singularities of the retarded kernels must be regularized by subtracting the corresponding singular $n$-point functions \cite{Detweiler:2002mi,Harte:2025tmd}.

\subsection{Equivalent Hamiltonian system}

In order to derive the Magnusian, it is convenient to switch to a Hamiltonian formulation of the dynamics. The dynamics derived from Eq. (\ref{eq:effectiveactionscalar}) can also be derived from the following action functional
\begin{equation}
    S[Q_+,Q_-]=S_0[Q_+,Q_-]-S_\text{nl}[Q_1,Q_2].
\end{equation}
Here, the unperturbed action is
\begin{equation}
    S_0[Q_+,Q_-]=\int\Big[ p_{+\mu} \dot{q}_{-}^\mu+p_{-\mu} \dot{q}_{+}^\mu-H_0(Q_+,Q_-)\Big]ds.
\end{equation}
The zeroth-order Hamiltonian is
\begin{equation}
    H_0(Q_+,Q_-)=Q_-^A \frac{\partial h(Q_+)}{\partial Q_+^A}
\end{equation}
where
\begin{equation}
    h(q,p)=-\sqrt{g^{\mu\nu}(q)p_\mu p_\nu}.
\end{equation}
The nonlocal piece is
\begin{equation}\label{eq:actioninpm}
    S_\text{nl}=\int dV_xdV_y \rho_-(x)G_\text{ret}(x,y)\rho_+(y)+\int dV_x dV_y dV_z \rho_-(x) J(x,y,z) \rho_+(y) \rho_+(z),
\end{equation}
up to higher-order corrections proportional to $\rho^3$. This action has the virtue that the roles of average and difference variables are clear. However, it still contains redundant information, since $\rho_-(x)$ is not linear in $q_-$. We can expand it as
\begin{eqnarray}
    \rho_-(x)&=&\rho_1(x)-\rho_2(x)\nonumber\\
    &=&k\int q_-^\mu (\tau) \partial_\mu \delta^{(4)}\big[x,q_+(\tau)\big]d\tau +O(q_-^2)
\end{eqnarray}
Hence, the nonlocal piece of the action functional can be expanded as
\begin{eqnarray}\label{eq:actioninpmexpanded}
    S_\text{nl}&=&-\int d\tau_1 d\tau_2 q_-^\mu(\tau_1) \partial_\mu^{(1)}G_\text{ret}\big[q_+(\tau_1),q_+(\tau_2)\big]\nonumber\\
    &-&\int d\tau_1 d\tau_2 d\tau_3 q_-^\mu(\tau_1) \partial_\mu^{(1)}J_\text{ret}\big[q_+(\tau_1),q_+(\tau_2),q_+(\tau_3)\big],
\end{eqnarray}
where $\partial^{(1)}_\mu$ is a derivative acting on the first entry of any n-point function.

The action functional in Eq.~\eqref{eq:actioninpmexpanded}, as written, cannot be used directly to derive the Magnusian. The reason is that it contains nonlocal-in-time interactions, due to the presence of n-point functions evaluated at different times. In Appendix \ref{appendixnonlocal}, we show how the nonlocalities can be treated perturbatively to produce a local Hamiltonian suitable for the treatment in Section~\ref{sec:magnusder}.

\subsection{Conservative/dissipative split}

In this subsection, we show that the in-in formalism is well-suited for defining the split between conservative and dissipative dynamics. First, we rewrite the action in the $1,2$ basis
\begin{eqnarray}
    S_\text{nl}&=&\frac{1}{2}\int dV_x dV_y G_C(x,y)\rho_1(x)\rho_1(y)-\frac{1}{2}\int dV_x dV_y G_C(x,y)\rho_2(x)\rho_2(y)\\
    &+&\int dV_x dV_y G_D(x,y)\rho_1(x)\rho_2(y)\nonumber\\
    &+&\frac{1}{4}\int dV_x dV_y dV_z J_C(x,y,z)\rho_1(x)\rho_1(y)\rho_1(z)-\frac{1}{4}\int dV_x dV_y dV_z J_C(x,y,z)\rho_2(x)\rho_2(y)\rho_2(z)\nonumber\\
    &+& \frac{3}{4}\int dV_x dV_y dV_z J_*(x,y,z)\rho_1(x)\rho_2(y)\rho_2(z)\nonumber\\
    &-&\frac{3}{4}\int dV_x dV_y dV_z J_*(x,y,z)\rho_2(x)\rho_1(y)\rho_1(z)\nonumber.
\end{eqnarray}
Here, we defined
\begin{subequations}
    \begin{eqnarray}
        G_C(x,y)&=&\frac{G_\text{ret}(x,y)+G_\text{ret}(y,x)}{2}\\
        G_D(x,y)&=&\frac{G_\text{ret}(x,y)-G_\text{ret}(y,x)}{2}\\
        J_C(x,y,z)&=&\frac{J_\text{ret}(x,y,z)+J_\text{ret}(y,x,z)+J_\text{ret}(z,x,y)}{3}\\
        J_*(x,y,z)&=&\frac{J_\text{ret}(x,y,z)-J_\text{ret}(y,x,z)-J_\text{ret}(z,x,y)}{3}
    \end{eqnarray}
\end{subequations}
This is not a satisfactory conservative/dissipative splitting yet. The reason is that we would like to inform the split by using the symmetry of the n-point functions under exchange of arguments. Following Ref.~\cite{Blanco:2026wwi}, we define the projector $\mathbb{P}_S$, which acts on functions $f(x_1,\dots,x_n)$ by symmetrizing under exchange of arguments
\begin{equation}
    \mathbb{P}_S f(x_1,\dots,x_n)=\frac{f(x_1,\dots,x_n)+\text{all permutations}}{n!}.
\end{equation}
In Ref. \cite{Blanco:2026wwi}, it was shown that the conservative and dissipative sectors can be defined using the projector $\mathbb{P}_S$. Namely, conservative $n$-point functions must be eigenfunctions of $\mathbb{P}_S$, while dissipative $n$-point functions must lie in its kernel. 

The 3-point function $J_*$ is not yet in the kernel of $\mathbb{P}_S$. This can be fixed by noting that there is freedom to redefine the action by terms of higher order in $\rho_-$. These terms do not affect the dynamics in the physical limit; they only represent a freedom to shuffle terms between what we define as conservative and dissipative. Hence, we add the term
\begin{equation}
    \frac{1}{12}\int dV_x dV_y dV_z J_C(x,y,z)\rho_-(x) \rho_-(y)\rho_-(z).
\end{equation}
This produces the equivalent nonlocal action
\begin{eqnarray}
    S_\text{nl}&=&\frac{1}{2}\int dV_x dV_y G_C(x,y)\rho_1(x)\rho_1(y)-\frac{1}{2}\int dV_x dV_y G_C(x,y)\rho_2(x)\rho_2(y)+\\
    &+&\int dV_x dV_y G_D(x,y)\rho_1(x)\rho_2(y)\nonumber\\
    &+&\frac{1}{3}\int dV_x dV_y dV_z J_C(x,y,z)\rho_1(x)\rho_1(y)\rho_1(z)-\frac{1}{3}\int dV_x dV_y dV_z J_C(x,y,z)\rho_2(x)\rho_2(y)\rho_2(z)\nonumber\\
    &+& \frac{1}{2}\int J_D(x,y,z)\rho_1(x)\rho_2(y)\rho_2(z)\nonumber-\frac{1}{2}\int J_D(x,y,z)\rho_2(x)\rho_1(y)\rho_1(z).
\end{eqnarray}
where the dissipative 3-point function is 
\begin{equation}
    J_D(x,y,z)=\frac{2J_\text{ret}(x,y,z)-J_\text{ret}(y,x,z)-J_\text{ret}(z,y,x)}{3}
\end{equation}
which now has no symmetric component. Note that setting all dissipative pieces to zero leads to two separate sectors: One that depends only on $\rho_1$ and one that depends only on $\rho_2$. Varying with respect to either path now leads to the correct equations of motion. In other words, the 2-point functions have a factor of $1/2$ and the 3-point functions have a factor of $1/3$. Hence, a variation of either trajectory will cancel these so that the equations of motion are proportional to the conservative $n$-point functions.

\section{Nonlocal forces and perturbative order reduction}\label{appendixnonlocal}

Integrating out mediating degrees of freedom generally produces effective forces that are nonlocal in time. This occurs, for example, in the gravitational self-force problem \cite{pound}, as well as at sufficiently high orders in the post-Newtonian and post-Minkowskian expansions \cite{Damour:2014jta,Dlapa_2022,PRLPorto4PMconservative,Bern_2022_PRL,Bern_2022_PoS}. In any of these approximations, nonlocal forces are generally expressed in terms of $n$-point functions
\begin{equation}
    F_A\big(Q_+\big)[Q_+]=\sum_{n=2}^\infty \epsilon^n \int_{-\infty}^{\infty} ds_2\dots ds_n \partial_A G_n\left[Q_+,Q_+(s_2),\dots,Q_+(s_n)\right],
\end{equation}
where the partial derivative acts only on the first entry of the n-point function. Even though the integration range extends to infinity, the causal support of the chosen propagators restricts the effective domain of integration. Here, $\epsilon$ denotes the small parameter controlling the perturbative expansion. Its precise meaning depends on the approximation under consideration; for example, $\epsilon= m/M$ in the self-force expansion, while in the post-Newtonian expansion it may represent powers of $v/c$.

In this appendix, we show how these nonlocal interactions can be systematically order reduced, yielding an equivalent local Hamiltonian description order by order in perturbation theory\footnote{This procedure is similar to the one developed in Ref. {\cite{blanco2024localhamiltoniandynamicsnonlocal}}. Both formalisms start with a doubling the degrees of freedom. However, Ref. \cite{blanco2024localhamiltoniandynamicsnonlocal} does not use the in-in-formalism. Instead, it is concerned with finding an equivalent Hamiltonian system within the original phase space describing only conservative dynamics. On the contrary, this paper never leaves the doubled phase space so the Hamiltonian obtained here includes dissipation.}.

\subsection{Doubled Hamiltonian for the order-reduced dynamics}

The dynamics of a nonlocal system can generally be derived from an action functional of the following form
\begin{equation}\label{eq:nonlocalaction}
\begin{aligned}
    \mathcal{S}&[Q_+,Q_-]=\int_{s_i}^{s_f}ds\Big[p_{+\mu}\dot{q}_-^\mu+p_{-\mu}\dot{q}_+^\mu-H_0(Q_+,Q_-)\Big]\\
    &-\sum_{n=2}^\infty \epsilon^n \int_{s_i}^{s_f}ds_1\int_{-\infty}^{\infty}ds_2\dots ds_n Q_-^A(s_1)\partial_A G_n\left[Q_+(s_1),Q_+(s_2),\dots,Q_+(s_n)\right].
\end{aligned}
\end{equation}
A variation with respect to $Q_-$ produces the equations of motion for the average variables
\begin{equation}\label{eq:euler+}
    E_A[Q_+]=\Omega_{AB}\dot{Q}_+^B-\frac{\partial H_0}{\partial Q_-^A}  -\sum_{n=2}^\infty \epsilon^n\int_{-\infty}^{\infty}ds_2\dots ds_n \partial_A G_n\left[Q_+,Q_+(s_2),\dots,Q_+(s_n)\right].
\end{equation}
This is an integro-differential equation for the path $Q_+(s)$. In general, the data required to specify the solution of an integro-differential equation is infinite dimensional. However, because the force is perturbative in $\epsilon$, the equations can be order reduced at each order. That is: At first order, the functional dependence on $Q_+(s)$ can be replaced by the zeroth-order solution $X^{(0)}$. Once the first-order solution is known, the second-order equations are solved by replacing the functional dependence with $X^{(1)}$ and so on. That means that, at least perturbatively, there is a unique path $X_{s_f,s_i}(Q)$ given a reference point $Q$, which satisfies Eq. (\ref{eq:euler+}). Hence, once the perturbative path $X$ has been specified, we may rewrite the force as
\begin{equation}
    F_A(Q_+,s)[X]=\sum_{n=2}^\infty \epsilon^n\int _{-\infty}^\infty ds_2 \dots ds_n \frac{\partial^{(1)}}{\partial Q_+^A}G_n \big[Q_+,X_{s_2,s}(Q_+),\dots,X_{s_n,s}(Q_+)\big].
\end{equation}
In this sense, the force $F_A$ is a local function of $Q_+$ once a path $X$ is specified. 

We can therefore replace Eq.~\eqref{eq:nonlocalaction} by the localized action
\begin{equation}\label{eq:localizedaction}
\begin{aligned}
    \mathcal{S}&[Q_+,Q_-;X]=\int_{s_i}^{s_f}ds\Big[p_{+\mu}\dot{q}_-^\mu+p_{-\mu}\dot{q}_+^\mu-H_0(Q_+,Q_-)\Big]\\
    &-\sum_{n=2}^\infty \epsilon^n \int_{s_i}^{s_f}ds_1\int_{-\infty}^{\infty}ds_2\dots ds_n Q_-^A(s_1)\partial_A G_n\left[Q_+(s_1),X_{s_2,s_1}\big(Q_+(s_1)\big),\dots,X_{s_n,s_1}\big(Q_+(s_1)\big)\right].
\end{aligned}
\end{equation}
This action is local in the sense that all instances of $Q_+$ and $Q_-$ are evaluated on the same time and, therefore, the Hamiltonian is a local function on the doubled phase space. When the solution $X$ is constructed perturbatively to order $\epsilon^N$, the action functionals in Eqs.~\eqref{eq:nonlocalaction} and \eqref{eq:localizedaction} produce the same equations of motion for $Q_+$ up to corrections of order $\epsilon^{N+1}$. However, these two action functionals will produce different equations for $Q_-$. This is not a problem since we do not require the localized functional to reproduce the full off-shell doubled dynamics of the original nonlocal action. We require only that it generate the order-reduced physical vector field and its associated tangent flow, since these determine the finite-time map on physical phase space. 

Since the action functional in Eq. (\ref{eq:localizedaction}) is strictly local in $Q_+$ and $Q_-$, it will be more convenient to use. In that case, its total variation is
\begin{equation}
\begin{aligned}
    \delta \mathcal{S}&=\int _{s_i}^{s_f}\left[\Omega_{AB}\dot{Q}_+^B-\frac{\partial H_0}{\partial Q_-^A}-F_A(Q_+,s)[X]\right]\delta Q_-^Ads\\
    &+\int _{s_i}^{s_f}\left[\Omega_{AB}\dot{Q}_-^B-\frac{\partial H_0}{\partial Q_+^A}-Q_-^C\frac{\partial  F_C(Q_+,s)[X]}{\partial Q_+^A}\right]\delta Q_+^Ads\\
    &+\left[p_{-\mu}\delta q_+^\mu +p_{+\mu}\delta q_-^\mu-H\delta s\right]^{s_f}_{s_i}
\end{aligned}
\end{equation}
The full Hamiltonian that appears as a boundary term is
\begin{equation}\label{eq:hamiltoniangalley}
\begin{aligned}
    H(Q_+,Q_-,s)[X]=&H_0(Q_+,Q_-)+Q_-^A\sum_{n=2}^\infty \epsilon^n \int_{-\infty}^\infty ds_2\dots ds_n \times \\\
   &\times \partial^{(1)}_AG_n\big[Q_+,X_{s_2,s}(Q_+),\dots,X_{s_n,s}(Q_+)\big].
\end{aligned}
\end{equation}
This Hamiltonian is local in $Q_+$ and $Q_-$ once the path $X$ has been determined perturbatively.

The Hamiltonian $H$ in Eq.~\eqref{eq:hamiltoniangalley} produces the equations of motion for $Q_+$ and $Q_-$ through
\begin{equation}
    \frac{dQ_\pm^A}{ds}=\{Q_\pm^A,H(Q_+,Q_-)[X]\}
\end{equation}

Finally, we note the force becomes time independent If the order-reduced flow $X$ is autonomous and the n-point functions are invariant under time translations. An autonomous flow is time-translation symmetric, meaning that
\begin{equation}
    X_{s,s_i}(Q)\equiv X_{s-s_i}(Q),
\end{equation}
In that case, a simple relabeling of the integration variables makes the force time-independent
\begin{equation}
    F_A(Q_+)[X]=\sum_{n=2}^\infty \epsilon^n\int _{-\infty}^\infty ds_2 \dots ds_n \frac{\partial^{(1)}}{\partial Q_+^A}G_n \big[Q_+,X_{s_2}(Q_+),\dots,X_{s_n}(Q_+)\big].
\end{equation}

\subsection{Interaction-Picture Magnusian} 

We now transform the localized interaction to the interaction picture. Let $X_{s,s_i}^{(0)}(Q) $ denote the unperturbed flow, and define its Jacobi propagator $M^A{}_B(s,s_i;Q)$ as in Eq.~(\ref{eq:defjacobi}).

Recall from Section \ref{sec:galley} that the difference variables evolve under the linearized flow
\begin{equation} 
Q_-^A(s) = M^A{}_B(s,s_i;Q_+) Q_-^B(s_i). 
\end{equation} 
where $M^A{}_B$ is the Jacobi propagator. The contraction of the Jacobi propagator and the partial derivative action on the first argument of the $n$-point function can be absorbed using the chain rule
\begin{equation}
M^B{}_A(s,s_i;Q) \partial_B^{(1)} G_n\left[ X_{s,s_i}^{(0)}(Q), \ldots \right] = \frac{\partial}{\partial Q^A} G_n\left[ X_{s,s_i}^{(0)}(Q), \ldots \right].
\end{equation} 
It is therefore convenient to define the interaction-picture $n$-point function by 
\begin{equation} 
\begin{aligned}
\mathcal{G}_n \left( Q_1,s_i,s_1; \ldots; Q_n,s_i,s_n  \right) &\equiv G_n\left[ X_{s_1,s_i}^{(0)}(Q_1), \ldots, X_{s_n,s_i}^{(0)}(Q_n) \right]. 
\end{aligned} 
\end{equation} 
Using the composition property of the unperturbed flow, the interaction-picture Hamiltonian becomes \begin{equation}\label{eq:nonlocalinteractionpicture} \begin{aligned} H^I(Q_+,Q_-,s,s_i) ={}& Q_-^A \sum_{n=2}^{\infty} \epsilon^{n} \int_{-\infty}^{\infty} ds_2\cdots ds_n \\ &\times \frac{\partial^{(1)}}{\partial Q_+^A} \mathcal{G}_n \left( Q_+,s_i,s; \ldots; Q_+,s_i,s_n  \right). 
\end{aligned}
\end{equation} 
Thus, the Jacobian associated with the unperturbed evolution of $Q_-$ is incorporated directly into the derivative of the interaction-picture n-point functions. 

The Hamiltonian in Eq. (\ref{eq:nonlocalinteractionpicture}) is a local function of $Q_+$ and $Q_-$, given that the zeroth order flow $X^{(0)}_{s_f,s_i}$ has been specified. It can therefore be inserted into the Magnus expansion derived in Section \ref{sec:magnusder}. To leading order in the interaction Hamiltonian, the nonlocal contribution to the Magnusian is
\begin{eqnarray}
\chi_{(1)}(Q_+,Q_-,s_f,s_i)=-\epsilon^2 Q_-^A  \int_{s_i}^{s_f} ds\int_{-\infty}^{\infty} ds_2 \frac{\partial^{(1)}}{\partial Q_+^A} \mathcal{G}_2 \left( Q_+,s_i,s, Q_+,s_i,s_2  \right). 
\end{eqnarray}
This is the nonlocal equivalent of Eq.~(\ref{eq:magnusiandefinition}). Higher-order corrections to the generator are obtained from the nested Poisson brackets of the corresponding interaction Hamiltonians.
	
\end{document}